\documentclass[amsmath,amssymb,11pt]{article}
\usepackage{jheppub2}
\pdfoutput=1
\usepackage{graphicx,epsfig}
\usepackage{float}
\usepackage{subfig}
\usepackage{subfloat}
\usepackage[utf8]{inputenc}
\usepackage{hyperref}
\usepackage{caption}
\usepackage{color}
\usepackage{overpic}
\usepackage[dvipsnames]{xcolor}
\usepackage{physics}
\usepackage[multiple]{footmisc}
\usepackage{comment}

\newcommand*{\affmark}[1][*]{\textsuperscript{#1}}

\newcommand{\beq}{\begin{equation}}
\newcommand{\eeq}{\end{equation}}

\newcommand{\gl}{\ell_\star}
\newcommand{\glbrane}{{\tilde \ell}_\star}

\newcommand{\gs}{g_\star}

\newcommand{\be}{\begin{equation}}
\newcommand{\ee}{\end{equation}}

\usepackage{tikz}
\usetikzlibrary{positioning}
\usetikzlibrary{intersections}
\usetikzlibrary{fadings} 
\usetikzlibrary{arrows.meta} 
\usetikzlibrary{arrows}

\tikzfading[name=fade out,
inner color=transparent!0,
outer color=transparent!100]

\definecolor{cherryblossompink}{rgb}{1.0, 0.72, 0.77}
\definecolor{lightblue}{rgb}{0.68, 0.85, 0.9}

\usetikzlibrary{decorations.pathmorphing}
\usetikzlibrary{decorations.pathreplacing,decorations.markings}

\usetikzlibrary{backgrounds,automata}

\title{Chemical Potential and Charge in Quantum Black Holes}

\author{Ana Climent,\affmark[1]}
\emailAdd{anacliment@icc.ub.edu}
\author{Roberto Emparan,\affmark[1,2]}
\emailAdd{emparan@ub.edu}
\author{and Robie A.~Hennigar\affmark[1]}
\emailAdd{robie.hennigar@icc.ub.edu}

\affiliation{
\affmark[1]Departament de Física Quàntica i Astrofísica and
  Institut de Ciències del Cosmos,\\
 Universitat de Barcelona, 08028 Barcelona, Spain\\
\affmark[2]Institució Catalana de Recerca i Estudis Avançats (ICREA)\\
 Passeig Lluis Companys, 23, 08010 Barcelona, Spain\\
}

\abstract{
We study systems in $2+1$ dimensions consisting of
defects that source an electric charge, or a magnetic flux, of a $U(1)$ field, and we use holography to compute their effects on quantum conformal fields. We can also hide the defects inside the horizon of a black hole, where they continue to affect the quantum fields outside. By extending the solutions to braneworld holography, we find the non-linear backreaction of the quantum fields on the defect and black hole backgrounds. This gives quantum charged point particles and black holes. The charged quantum black holes markedly differ from classically charged BTZ black holes, since the quantum-induced electromagnetic field in $2+1$ dimensions has a better asymptotic behavior than its classical counterpart. The construction also gives a new class of (near-)extremal charged quantum black holes with AdS$_2$ throats.
}

\begin{document}

\maketitle

\section{Introduction}

Gravitational physics and condensed matter theory provide good motivation to investigate strongly coupled quantum theories in the presence of chemical potentials and curved background geometries. In this article, we will study systems in $2+1$ dimensions consisting of a defect that sources an electric charge, or a magnetic flux, of a $U(1)$ field. In fact, we can consider dyonic defects that carry both electric charge and magnetic flux (which may provide a model for anyons). We will find the effects induced by these defects on quantum conformal fields. Perhaps more intriguingly, we can also hide the defects inside the horizon of a black hole, where they will continue to have an effect on the quantum conformal fields outside. The study of all these systems through conventional quantum field theory methods can often be difficult and of limited reach. However, we will see that holography provides an efficient means to solve the problem.

Viewed another way, we are dressing the defects and the black holes with quantum conformal fields that get excited, or polarized, by the background where they live. An interesting question is how these quantum fields affect the backgrounds when their backreaction---gravitational or electromagnetic---is large enough that it must be taken into account. Holography, in its version as `braneworld holography', allows us to compute this backreaction. For a range of parameters (charge smaller than the mass---we will make this precise), the gravitational effect of quantum fields gives rise to quantum black holes which are horizon-dressed conical defects or quantum-corrected black holes. These were explored, as neutral solutions, in \cite{Emparan:2002px,quBTZ}. Here, with the addition of a chemical potential that becomes a dynamical gauge field, we will see how quantum black holes charge up in a manner that differs markedly from the classically charged black holes. In particular, the quantum-induced electromagnetic field in $2+1$ dimensions will have better asymptotic behavior than its classical counterpart. A notable outcome is a new class of (near-)extremal charged black hole solutions.

\paragraph{Chemical defects: charged, conical, and hidden.}

The simplest system that we study consists of a CFT in the presence of a defect which sources a chemical potential of the form
\begin{align}\label{bchempot}
    A=\frac{a}{r}\,dt+\mu_\textrm{m}\, d\phi
\end{align}
at the origin of flat $2+1$ spacetime
\begin{align}\label{M3}
    ds^2=-dt^2+dr^2 +r^2 d\phi^2\,.
\end{align}
This potential, with constant $a$ and $\mu_{m}$, introduces a marginal deformation in the CFT and corresponds to a pointlike electric charge and magnetic vortex. We will solve the state of the CFT using holography, namely, by finding a dual bulk solution of the Einstein-Maxwell-AdS$_4$ theory whose potential and geometry at the boundary are \eqref{bchempot} and \eqref{M3}. 

This is not a new problem: the solution for a purely electric defect ($a\neq 0$, $\mu_\textrm{m}=0$) was found and analyzed in \cite{Horowitz:2014gva}.\footnote{See also~\cite{Lu:2014sza} for related considerations.} Adding the magnetic flux $\mu_\textrm{m}$ is the first of the extensions that we will investigate in this article. We will then consider the defect in non-trivial geometries: first by coupling it to a conical defect in the planar geometry, and then by putting it in a negatively curved spacetime, namely AdS$_3$, with or without a conical defect. That is, instead of \eqref{M3} we consider the two-parameter family of spacetimes
\begin{align}\label{AdS3}
    ds^2=-\left(\alpha^2+\frac{r^2}{\ell_3^2}\right)dt^2+\frac{dr^2}{\alpha^2+\frac{r^2}{\ell_3^2}} +r^2 d\phi^2\,.
\end{align}
When $\ell_3\to\infty$ and $\alpha\neq 1$, this is a locally flat spacetime with a conical deficit angle (which is also a marginal deformation)
\begin{align}\label{def}
    \delta=2\pi\left(1-\alpha\right)
\end{align}
at $r=0$. Such geometries may be engineered in lab systems but are also of general interest in gravitational theory, e.g., as sections of cosmic strings. When $\ell_3<\infty$, the defect (with the same deficit angle \eqref{def}) lives in an AdS$_3$ spacetime of locally constant negative curvature with radius $\ell_3$. This curvature acts as a potential well that pushes the field energy towards the origin. 

It is now a simple step to extend the scheme and hide the defect inside a black hole. For this purpose, we replace \eqref{AdS3} with the BTZ black hole geometry \cite{Banados:1992wn,Banados:1992gq}
\begin{align}\label{BTZ}
    ds^2=-\left(\frac{r^2-r_+^2}{\ell_3^2}\right)dt^2+\ell_3^2\frac{dr^2}{r^2-r_+^2} +r^2 d\phi^2\,,
\end{align}
which has a black hole horizon at $r=r_+$. In this case, the chemical potential can only be regular if, in addition to  \eqref{bchempot}, there is a constant, relevant term 
\begin{align}
    \mu_e=-\frac{a}{r_+} 
\end{align}
for the electric component, such that the potential
\begin{align}
     A=\left(\mu_e+\frac{a}{r}\right)\,dt
\end{align}
vanishes on the horizon. 

In this article we will describe the states of a holographic CFT in all these backgrounds by constructing four-dimensional bulk spacetimes that are dual to them. Using them, we will compute the expectation values of the stress-energy tensor and the $U(1)$ current induced by the defect. In contrast with \cite{Horowitz:2014gva}, we will argue that these defects must not be assigned an entropy.

\paragraph{Charged quantum black holes and point defects in $2+1$ dimensions.}

At this point, one may notice an apparent oddity, especially in the case of a black hole in a chemical potential. The $U(1)$ symmetry in this problem is not dynamical. If this meant that it is a global symmetry, then the charge of the defect would be a global charge, and so would, too, the charge of the black hole. However, black holes are notorious for not supporting global charges. 

In the setup above, the catch is readily apparent. The $U(1)$ symmetry is `weakly gauged', namely, the CFT couples to the gauge potential $A$ (think of the gauge-covariant derivative) but the dynamical term $\propto F^2$ for the gauge field is suppressed. The weak gauging allows to compute the vev of the CFT currents that a dynamical gauge field would couple to\footnote{This is analogous to computing the stress tensor in a non-gravitational theory from the variation of the action with respect to the metric. There, we weakly gauge the diff symmetry to compute the stress tensor that dynamical gravity would couple to.}, and therefore the black hole can be charged.

For this to be consistent, the black hole itself must not be dynamical, and indeed the boundary spacetimes \eqref{AdS3} and \eqref{BTZ}, like the gauge fields \eqref{bchempot}, are fixed backgrounds for the CFT. An extension of holographic duality, namely braneworld holography, allows us to switch on dynamics for both gravity and the gauge field.

By doing this, braneworld holography automatically includes the backreaction of the quantum CFT on the geometries \eqref{AdS3} and \eqref{BTZ} and on the potential \eqref{bchempot}. In the absence of charge, it was discovered in \cite{Emparan:2002px,quBTZ} that this quantum backreaction has notable consequences. First, it dresses the conical singularities with horizons, to yield quantum black holes in regimes of $2+1$ gravity where otherwise there would not be any classical black holes. Second, it allows the incorporation of the non-linear quantum backreaction on the BTZ black hole. This is the quantum BTZ black hole.\footnote{See \cite{Casals:2016odj} for the backreaction of free fields on the BTZ black hole.} 

In this article, we will draw on the methods of \cite{Emparan:2002px,quBTZ} to construct novel charged quantum black holes in 2+1 dimensions, and also solutions for charged point sources that backreact on the geometry without creating a horizon.

\emph{Charged quantum black holes.} The solution of a classical charged BTZ black hole coupled to a Maxwell field was obtained in \cite{Banados:1992wn} (see also \cite{Peldan:1992mp,Clement:1995zt,Martinez:1999qi}), but the logarithmic radial dependence of the classical gauge field in $2+1$ dimensions endows it with rather pathological features. Our solution is different in crucial respects. The dynamics of the gauge field is induced by quantum effects, namely by integrating out the CFT degrees of freedom above a UV cutoff. This is the same phenomenon that induces gravitational dynamics in the $2+1$ system, and it results in a theory of gravity and electromagnetism with a tower of higher derivative corrections for both fields. The fact that gauge field dynamics is induced by loop effects and involves a specific infinite tower of higher-order terms (e.g., $RF^2$, $\nabla F\nabla F$, etc), implies that the classical $\ln r$ dependence is absent, and is instead replaced by a $1/r$ fall-off. As a result, the asymptotic behavior of the solution poses no difficulties.

\emph{Charged point sources.} When charge is present the backreaction will not dress all the defects with a horizon. If the charge to mass ratio of the defect is large enough, we find backreacted charged point sources that have naked curvature singularities. These solutions should not be discarded away, for the same reason that the Reissner-Nordström solutions with $Q>M>0$ are not regarded as pathological. They correctly describe the gravitational field created by pointlike charged sources down to the distance where quantum electrodynamics modifies the electromagnetic field around the charge. This occurs a Compton wavelength away from the source, and farther than this the classical solution remains valid. In contrast, in the absence of charge the solutions with negative energy are as pathological as the negative-mass Schwarzschild solution. All the conical defects with positive energy above the ground state develop a horizon around them.

The space of solutions of charged quantum black holes is very rich, and we will only begin to uncover their thermodynamics (in this respect, see \cite{Frassino:2023wpc} for the possibilities that quantum black holes afford). Their causal structure is like that of the Reissner-Nordstr{\"o}m solution, with an inner horizon that, arguing like \cite{Emparan:2020rnp,Kolanowski:2023hvh}, is expected to become singular at the next order in quantum backreaction. 

\paragraph{Near-extremal quantum throats.} A particularly interesting feature is that the quantum charged black holes have extremal and near-extremal regimes where they develop long AdS$_2$ throats. Recently, the dynamics of black holes near extremality has been shown to be dominated by \emph{quantum gravitational} fluctuations of the throat \cite{Iliesiu:2020qvm}. These are not part of the quantum effects in our solution, which come only from the CFT and are an $O(1/G)$ effect from the large number of holographic CFT degrees of freedom. In the region far from the throat, these effects are much larger than the quantum fluctuations of the geometry, but the latter dominate in the throat if the temperature is sufficiently low. Related considerations have appeared earlier in \cite{Kolanowski:2023hvh}. Although in this article we will not study the Schwarzian theory of the throat, we will obtain the parameters and main features of the near-extremal solutions.

These new extremal black holes in 2+1 dimensions necessitate that the charge backreaction on the horizon is comparable to chargeless effects. Thus they need a nonzero (i.e., not perturbative) backreaction, and when this is small there are two classes of solutions: black holes whose horizons result from the quantum backreaction on a charged horizonless defect, or small-mass BTZ black holes (with small area) that become extremal through the addition of quantum charge.

The quantum extremal black holes appear at the parameter boundary between the charged quantum black holes and the point sources, but there is no discontinuous jump in the entropy when crossing this boundary. Since the quantum fluctuations of the AdS$_2$ throat bring down the degeneracy of the extremal black hole, this describes a single microscopic state \cite{Iliesiu:2020qvm}. In this sense, the extremal black hole is more like the charged point defects, with the geometry dominated by large quantum fluctuations. 

More recently, it has also been realized that when extremal black holes are deformed away from spherical shape, their horizon generically develops a tidal singularity \cite{Horowitz:2022mly,Horowitz:2024dch}. The horizon of our extremal black holes is not spherical but it is nevertheless smooth because the solutions are type D and the Weyl tensor is algebraically special. A more generic deformation is expected to turn them singular.

\bigskip The plan for the remainder of the article is as follows. In section~\ref{sec:DoubleWicks} we construct and analyze the holographic bulk duals of the quantum conformal fields in the fixed geometric and chemical backgrounds described above. We obtain them through double-Wick rotation of Reissner-Nordström-AdS$_4$ solutions, extending the constructions in \cite{Hubeny:2009rc,Horowitz:2014gva}. In section~\ref{sec:ChqBTZ} we use braneworld holography as in \cite{Emparan:2002px,Emparan:2022ijy} to compute the backreaction of the quantum conformal fields on the geometry and the gauge field, and examine the properties of the new solutions, in particular of the new quantum charged BTZ black holes and their extremal limits. We then conclude in section~\ref{sec:discuss}. To improve the readability of  sections~\ref{sec:DoubleWicks} and \ref{sec:ChqBTZ} without omitting relevant detail, many of the technical analyses are postponed to a series of appendices.

\paragraph{N.B.} While we were writing up this article, ref.~\cite{Feng:2024uia} appeared in arXiv, which overlaps with our Section~\ref{sec:ChqBTZ}.

\section{Holographic CFT on Curved Spacetime with a Chemical Potential}
\label{sec:DoubleWicks}

We will now construct the bulk solutions that describe the coupling of a holographic CFT to the chemical defects \eqref{bchempot} and geometries \eqref{AdS3} at the boundary, and from these solutions we extract the relevant CFT data. In all cases, these are solutions of Einstein-Maxwell theory
\be \label{bulk_action}
I = \frac{1}{16 \pi G_4} \int d^4x \sqrt{-g} \left[\frac{6}{\ell_4^2} + R - \dfrac{\gl^2}{4} F^2 \right] \, .
\ee
We measure the coupling between the Einstein and Maxwell terms using a length scale $\gl$. Equivalently, this can be expressed in terms of the dimensionless gauge coupling constant $\gs$ by
\be 
\gl^2 = \frac{16 \pi G_4}{\gs^2} \, .
\ee

The solutions of this theory are holographic duals to states of a 2+1 dimensional CFT with a chemical potential. Its central charge is
\begin{align}\label{central}
    c=\frac{\ell_4^2}{G_4}
\end{align}
up to a factor that depends on the concrete realization of the duality. Following \cite{Emparan:2020rnp}, we will keep it simple and fix this factor as in \eqref{central} without being more specific about it. Observe that the 2+1-dimensional theory only specifies dimensionless ratios of 3+1 quantities, namely
\begin{align}
    \frac{\ell_4^2}{\gl^2}=\frac{\gs^2 c}{16\pi}\,.
\end{align}

\subsection{Dyonic Defects in (conical) AdS$_3$ and Minkowski}\label{condefs}

Equations \eqref{bchempot} and \eqref{AdS3}
describe a dyonic defect in a conical AdS$_3$ (for finite $\ell_3$) or conical Minkowski geometry (for $\ell_3 \to \infty$). 
Only in the case $\alpha=1$ is the geometry global AdS$_3$ or Minkowski.

\paragraph{Bulk solution.} An exact solution of the theory \eqref{bulk_action} with these boundary data is 
\begin{align}\label{dyonic_defect_ads}
ds^2 &= \frac{\rho^2 \alpha^2}{r^2} \left[- \left(\alpha^2 + \frac{r^2}{\ell_3^2} \right) dt^2 +  \frac{dr^2}{\alpha^2 + \frac{r^2}{\ell_3^2}}  + \frac{r^2 \ell_4^2}{\rho^2} F_{-1}(\rho) d\phi^2\right] + \frac{d\rho^2}{F_{-1}(\rho)}\,,
\\ 
A &= \frac{a}{r} dt + \mu_\textrm{m} \left(1-\frac{\rho_0}{\rho} \right) d\phi \, ,
\end{align}
where 
\be\label{4dmet} 
F_{-1}(\rho) = \frac{\rho^2}{\ell_4^2} - 1 - \frac{2 G_4 m}{\rho} - \frac{Q^2}{\rho^2} \,.
\ee
Here $m$ and $Q$ are integration constants of the bulk solution. The constant $Q$ is related to the parameters appearing in the boundary chemical potential according to
\be\label{Qrel} 
    Q^2 = \frac{\gl^2}{4 \alpha^2} \left[ \frac{a^2}{\alpha^2} + \frac{\rho_0^2 \mu_\textrm{m}^2}{\ell_4^2} \right] \, ,
\ee
and the interpretation of $\rho_0$  will be clarified momentarily. A detailed analysis of the solution is provided in Appendix~\ref{App_DW}, where we show that it is equivalent to a double Wick rotation of the familiar Reissner-Nordstr{\"o}m AdS black hole with hyperbolic transverse geometry. Here we summarize the salient features.  

The physically sensible solutions are those for which the bulk metric function $F_{-1}(\rho)$ has a zero at $\rho = \rho_0>0$, i.e.,
\begin{align}\label{Frho0}
    F_{-1}(\rho_0)=0\,,
\end{align}
since otherwise the diverging curvature at $\rho\to 0$ is naked. Then the action of $\partial_\phi$ has a fixed point at $\rho=\rho_0$, and for generic parameter values the solutions exhibit conical singularities in the $(\rho, \phi)$ section of the metric. Eliminating them requires that
\be \label{alphaeq}
\alpha = \frac{2}{\ell_4} \frac{1}{| F'_{-1}(\rho_0) |} \,.
\ee
With this identification, the bulk geometry is completely regular. While the boundary will generically have a conical singularity at $r = 0$ this feature does not extend into the bulk, as a consequence of the conformal factor. 

\paragraph{CFT state.} Equations \eqref{Qrel}, \eqref{Frho0} and \eqref{alphaeq} give us the complete solution to the state of the holographic CFT in the presence of the background potentials \eqref{bchempot} and geometry \eqref{AdS3}. Namely, once we fix the boundary data $a$,  $\mu_\textrm{m}$ and $\alpha$, these three equations determine $m$, $Q$ and $\rho_0$ in terms of them so that the bulk solution is completely specified. Note that $\ell_3$ is not involved since it does not appear in \eqref{Qrel}, \eqref{Frho0} and \eqref{alphaeq}, so the relationships are between the dimensionless parameters $(\alpha, a, \mu_\textrm{m})$ and $(m/\ell_4,Q/\ell_4,\rho_0/\ell_4)$.
These have been worked out in appendix~\ref{App_alpha}. In addition, recall that the parameters of the Einstein-Maxwell theory $G_4$, $\ell_4$ determine the central charge $c$ of the CFT, \eqref{central}, while $\gl$ controls how the CFT couples to the chemical potential. With this, any quantity computed using the bulk solution can be translated into a property of the CFT state.

It is now straightforward to obtain the entries of the holographic dictionary, e.g., by transforming the metric to Fefferman-Graham gauge. The azimuthal component of $A$ sources a non-trivial response of the CFT, which along with the stress-tensor read
\begin{align} \label{TmunuDWAdS} 
    \langle T_\mu^\nu \rangle  &=  \frac{3 c}{16 \pi } \frac{2 m \alpha^3}{3 \ell_4 r^3} \, {\rm diag} \left\{1,1, -2 \right\}  \, , 
    \\ 
    \langle  J^{\phi} \rangle & = \dfrac{\mu_\textrm{m} \alpha \rho_0}{\ell_4 r^3}  \, . \label{JmuDWAdS}
\end{align}
According to the discussion above, here $m$ and $\rho_0$ must be regarded as functions of the CFT background parameters $\alpha$, $a$ and $\mu_\textrm{m}$. Any possible dependence of  \eqref{TmunuDWAdS} and \eqref{JmuDWAdS} on $\ell_3$ is restricted by conformal invariance, since on dimensional grounds the radial dependence must be of the form $r^{-3} f(r/\ell_3)$. That the holographic solution has constant $f(r/\ell_3)$, and hence no dependence on $\ell_3$, has been argued in \cite{Emparan:2002px,Emparan:2020rnp} to be a consequence of the transparent boundary conditions for the conformal fields that the holographic construction implicitly imposes. The argument in \cite{Emparan:2002px,Emparan:2020rnp} was made for neutral systems, but we expect it is also valid for the charged systems.

The bulk field strength has non-trivial components tangent to the boundary, 
\be \label{Ftr_bdry}
F^{tr}|_{\rm bdry} =  - \dfrac{a}{ r^2} \, .
\ee
This should be interpreted to mean that the CFT is immersed in a background electric field of precisely this character.  This situation is analogous to the dyonic AdS$_4$ black hole, for which the dual description is a CFT at finite chemical potential immersed in a background magnetic field~\cite{Hartnoll:2009sz}.

The boundary contains a point charge defect localized at $r = 0$. This can be seen by considering the computation of the charge of the bulk solution. Let us change $r/\alpha =\alpha/\tilde{r}$ in \eqref{dyonic_defect_ads} so the metric becomes
\be \label{near_defect_ads}
    ds^2 = \rho^2 \left[-\left(\tilde{r}^2+\frac{\alpha^2}{\ell_3^2}\right) dt^2 +  \frac{d\tilde{r}^2}{\tilde{r}^2+\frac{\alpha^2}{\ell_3^2}}\right] + \alpha^2 \ell_4^2F_{-1}(\rho) d\phi^2 + \frac{d\rho^2}{F_{-1}(\rho)}\,.
\ee
In this form it is apparent that the $(t,\tilde{r})$ part of the geometry describes an AdS$_2$ spacetime, with the location $r=0$ of the defect corresponding to the AdS$_2$ boundary at $\tilde{r}\to\infty$. 
The charge of the defect may be computed by integrating the electromagnetic flux at this boundary, with the result that
\be
    Q_e = \dfrac{1}{g_\star^2 } \int \star F = \dfrac{2 \pi a }{\alpha g_\star^2} \, , \qquad Q_{\rm m} = \dfrac{1}{g_\star^2 } \int F = \frac{2 \pi  \mu_\textrm{m} \rho_0 }{\ell_4 g_\star^2} \, .
\ee
These charges are bulk quantities and thus they give the expected interpretation of the total charge \eqref{Qrel} as
\be
    Q^2 = \dfrac{g_\star^2 G_4}{\pi \alpha^2} \left( Q_e^2 + Q_{\rm m}^2 \right) \,.
\ee
Moreover, they can be regarded as parameters of the CFT  with a weakly gauged $U(1)$ symmetry, allowing to write
\be 
\langle  J^{\phi} \rangle = \frac{\alpha}{2\pi} \frac{\gs^2 Q_{\rm m}}{r^3}\,.
\ee

\paragraph{No defect entropy.} The authors of \cite{Horowitz:2014gva} noted that, in the case $\ell_3^2\to\infty$ that they studied, the AdS$_2$ geometry has an extremal horizon at $\tilde{r}=0$ ($r\to\infty$). They interpreted the (regularized) area of this horizon as a defect entropy.  However, this does not seem appropriate for two reasons. First, recent studies have shown that the entropy of extremal non-supersymmetric horizons vanishes once the quantum fluctuations of the AdS$_2$ throat are included \cite{Iliesiu:2020qvm}. Here we will not perform this analysis (which should be similar to the one in \cite{Emparan:2023ypa}) but we expect that the same result holds. Second, the extremal horizon is absent when $\ell_3<\infty$, which corresponds to defects in AdS$_3$ instead of flat spacetime. It would seem odd that an intrinsic property of the defect such as its entropy should depend on the asymptotics of the spacetime it resides in. The vanishing of the extremal horizon entropy is consistent with this view.

\begin{figure}[t]
\centering
\includegraphics[width=0.65\textwidth]{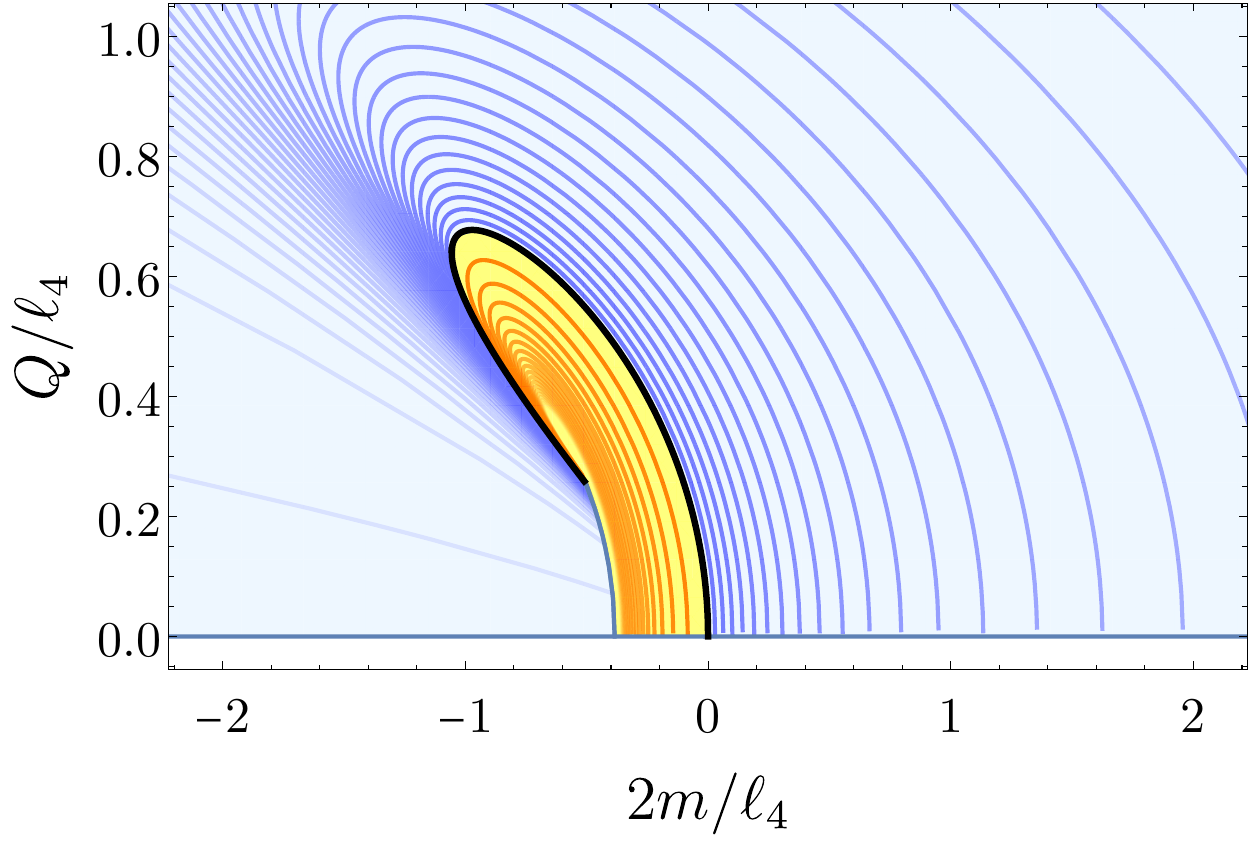}
\caption{\small The conical deficit/excess is shown across the $(m, Q)$ parameter space, where, roughly, $Q$ gives a measure of the defect charge and $m$ the stress-energy of the quantum fields. The blue-shaded region corresponds to conical deficits, while the yellow-shaded region corresponds to conical excesses. The solid, black curve corresponds to the family of smooth geometries with $\alpha = 1$. Superposed on this plot we show several contours of constant $\alpha$. The blue curves correspond to conical defects with constant $\alpha < 1$ and the red curves correspond to conical excesses with constant $\alpha > 1$. The opacity of the curves decreases away from the regular solution with lighter colors corresponding to a greater defect (either excess or deficit).}
\label{defect_phase_space}
\end{figure}

\paragraph{Parameter space.} A detailed analysis of the allowed deficit angle is performed in appendix~\ref{App_alpha}, with the final results displayed in Figure~\ref{defect_phase_space}. It shows the conical deficit/excess over the $(m, Q)$ parameter space.\footnote{Here $Q$ is the parameter appearing in the metric function~\eqref{4dmet}, meaning both electric and magnetic cases are included in the analysis.} It is possible to have both conical deficits and excesses (with $\alpha\lessgtr 1$), along with a one-parameter family of smooth boundary geometries ($\alpha = 1$). When the chemical potential vanishes, the only case with a regular boundary geometry is when the bulk is AdS$_4$ and the boundary is AdS$_3$. With chemical potential, the situation is actually much more intricate, with regular geometries existing over a range of $m$ and $Q$ parameters.  All of these states, except for the one with $m = 0$, have negative energy density, which must be regarded as a Casimir energy.

Finally, let us highlight that for a certain range of parameters the plot is double-valued. In these cases, the boundary geometry with a given defect charge has two corresponding bulk duals. That is, for certain values of the geometric and chemical deformations, there are two different CFT states. These two possibilities are distinguished by the stress-tensor, which is sensitive to the value of $m$. It is plausible that the state with lowest energy is preferred.

\subsection{Dyonic Defects inside BTZ Black Holes}\label{BTZdefs}

It is also possible to hide the charged defect inside a horizon. We therefore consider the following boundary geometry and chemical potential,
\begin{align}\label{BTZ_bdry}
    ds_{\rm bdry}^2 =&\, -f(r) dt^2 + \frac{dr^2}{f(r)} + r^2 d\phi^2 \,  \quad \text{with} \quad f(r) = \frac{r^2-r_+^2}{\ell_3^2} \, ,
    \\ 
    A_{\rm bdry} =& \left(\mu_e + \frac{a}{r} \right) dt + \mu_\textrm{m} d\phi \, .
\end{align} 
The Einstein-Maxwell equations admit an exact solution with these boundary data at $\rho\to\infty$, namely,
\begin{align}
    ds^2 &= \frac{\rho^2 r_+^2}{r^2 \ell_3^2} \left[- \frac{r^2-r_+^2}{\ell_3^2} dt^2 + \frac{\ell_3^2}{r^2-r_+^2} dr^2 + r^2 \frac{\ell_4^2 F_{+1}(\rho)}{\rho^2} d\phi^2 \right] + \frac{d\rho^2}{F_{+1}(\rho)} \, , \label{DWRN_BTZ}
    \\
    A &=  \left(\mu_e + \frac{a}{r}\right) dt + \mu_\textrm{m} \left(1 - \frac{\rho_0}{\rho} \right) d\phi  \,  \label{DWRN_BTZ_gauge_field} \,,
\end{align}
with
\be 
    F_{+1} = \frac{\rho^2}{\ell_4^2} + 1 - \frac{2 G_4 m}{\rho} - \frac{Q^2}{\rho^2}
\ee
and $Q$ now takes the form 
\be\label{QrelBTZ} 
    Q^2 = \frac{\gl^2 \ell_3^2}{4 r_+^2} \left[ \frac{a^2\ell_3^2}{r_+^2} + \frac{\rho_0^2 \mu_\textrm{m}^2}{\ell_4^2} \right] \, .
\ee
Analogous to the conical AdS$_3$/Minkowski geometries, this bulk solution can be obtained as a double Wick rotation of the spherical AdS$_4$ Reissner-Nordstr{\"o}m black hole.\footnote{This is an extension of the construction in \cite{Hubeny:2009rc} to include charge.} The analysis is very similar to the previous one for the defects, so we will proceed more quickly, only emphasizing the differences that the horizon introduces.

The new feature here is the appearance of $\mu_e$ in the gauge potential. This is necessary to have a regular gauge field at the horizon at $r = r_+$, which fixes it to
\begin{align}
    \mu_e = -\frac{a}{r_+}\,. 
\end{align}
Therefore $\mu_e$ must not be thought of as an independent parameter in the boundary data.

As before, the relevant bulk geometries are those for which the metric function $F_{+1}(\rho)$ vanishes at $\rho_0 > 0$. This leads to a regularity requirement in the bulk that fixes the horizon radius of the boundary black hole in terms of bulk quantities,
\be\label{BTZrad} 
    r_+ = \dfrac{\ell_3}{\ell_4} \frac{2}{|F_{+1}'(\rho_0)|} \, .
\ee
Conversely, for the CFT interpretation we regard \eqref{BTZrad} together with \eqref{QrelBTZ} and $F_{+1}(\rho_0)=0$ as determining the bulk solution parameters $(m, Q, \rho_0)$ (in units of $\ell_4$) in terms of the CFT boundary data $(r_+,a,\mu_\textrm{m})$. The relationships are given in appendix~\ref{app:BTZ_params}. 
The bulk solution then provides the state of a strongly coupled CFT deformed by the chemical potential above, in a thermal equilibrium state on a non-dynamical BTZ black hole. The horizon radius directly determines the temperature of the boundary system,
\be \label{TBTZ}
T = \frac{r_+}{2\pi \ell_3^2} \, .
\ee

A simple analysis of the polynomial $F_{+1}(\rho)$ allows one to make some useful conclusions about this system. First, by applying Descartes' rule of signs to the polynomial it can be determined that there is always a single real, positive root $\rho_0$. A more detailed analysis shows that at fixed $Q$, $F'_{+1}(\rho_0)$ always has a minimum as a function of $\rho_0$. This translates into an upper bound on the temperature of the boundary BTZ black hole. The upper bound is $Q$-dependent with an inverse relationship. That is, for larger values of $Q$ the maximum temperature of the boundary black hole is lower. The global maximum of temperature is achieved in the case $Q = 0$ and decreases to zero as $Q \to \infty$. 

The standard tools of holographic renormalization give the CFT data
\begin{align} \label{TmunuDWBTZ} 
    \langle T_\mu^\nu \rangle  &=  \frac{3 c }{16 \pi} \frac{2 m r_+^3}{3 \ell_4 \ell_3^3 r^3} \, {\rm diag} \left\{1,1, -2 \right\}  \, ,
    \\ \label{JmuDWBTZ}
    \langle J^{\phi} \rangle  &= \frac{\mu_\textrm{m} \rho_0 r_+}{\ell_4 r^3\ell_3} \, ,
\end{align} 
and the same background electric field at the boundary as before, \eqref{Ftr_bdry}. For the same reasons as before, all the dependence on $r_+$ and $\ell_3$ enters through their ratio $r_+/\ell_3$.

There is a defect at the boundary at $r = 0$, but now it is inside the horizon of the black hole. For the BTZ black hole \eqref{Ftr_bdry}, this is a null singularity. Computations of the charge can be carried out in the same manner as before to find
\be 
 Q_e = \dfrac{2 \pi a \ell_3 }{\gs^2 r_+} \, , \qquad Q_{\rm m} = \dfrac{2 \pi   \mu_\textrm{m} \rho_0}{\gs^2 \ell_4 } \, .
\ee
Then
\be\label{Q2BTZ}
    Q^2 = \dfrac{g_\star^2 G_4 \ell_3^2}{\pi r_+^2} \left( Q_e^2 + Q_{\rm m}^2 \right) \,,
\ee
and \eqref{JmuDWBTZ} can be expressed as
\be \label{JQ}
\langle  J^{\phi} \rangle = \frac{r_+}{2\pi \ell_3} \frac{\gs^2 Q_{\rm m}}{r^3}\,.
\ee

The defect is hidden inside a black hole horizon, but since the boundary geometry is not dynamical, it is not appropriate to view the horizon area as giving the entropy of the BTZ black hole at the boundary. Instead, it must be interpreted as the entanglement entropy of the CFT across this horizon. This is infinite since the horizon extends all the way out to the boundary at $\rho\to\infty$, but it would be possible to compute finite entropy differences between states with the same boundary geometry, i.e., the same value of $r_+$ but different values of $a$ and $\mu_\textrm{m}$. We expect that the entanglement entropy decreases as the potential increases.

\section{Quantum Backreaction on Charged Defects and Black Holes }\label{sec:ChqBTZ}

To understand how the charged CFT states backreact on the geometry and the gauge field, we shall use the approach of braneworld holography, which provides a solution to the effective Einstein-Maxwell theory with higher-derivative terms for both the geometry and the gauge field. Specifically, we are interested in the charged version of the quBTZ black hole \cite{quBTZ}. This is possible, with remarkable analytic control, due to the existence of a four-dimensional solution of the Einstein-Maxwell-AdS theory that describes a three-dimensional quantum black hole on a brane, namely, the AdS$_4$ C-metric, which has a dyonic extension including electric and magnetic charges of the bulk black hole \cite{Plebanski:1976gy}. The metric can be conveniently written in the form
\beq \label{cmetric}
    ds^2 = \dfrac{\ell^2}{(\ell+xr)^2}\left[ -H(r)dt^2 + \dfrac{dr^2}{H(r)} + r^2 \left( \dfrac{dx^2}{G(x)} + G(x)d\phi^2 \right) \right],
\eeq
with
\beq \label{Hr}
    H(r) = \dfrac{r^2}{\ell_3^2} + \kappa - \dfrac{\mu \ell }{r} + \dfrac{q^2 \ell^2}{r^2} \, ,  \qquad  G(x) = 1-\kappa x^2 - \mu x^3 - q^2x^4  \, .
\eeq 
where the parameters $\ell_3$, $\kappa$, $\mu$, $q$ and $\ell$ will be soon clarified. Following the same conventions as in the previous section, the gauge field is taken to be 
\beq \label{gauge_field_bulk}
    A =  \dfrac{2 \ell }{\gl} \left[  e \left(\frac{1}{r_+} - \frac{1}{r} \right) dt + g \left(x - x_1 \right) d\phi \right] \, , 
\eeq
where $e$ and $g$ are related to the electric and magnetic charge of the black hole,
\be 
q^2 = e^2 + g^2 \, .
\ee
Note that we have chosen a gauge for the potential that is regular at the relevant zeroes of $H(r)$ and $G(x)$, namely $H(r_+)=0$ and $G(x_1)=0$, where $r_+$ is the largest positive root and $x_1$ is  smallest positive root of the respective polynomial. To ensure Lorentzian signature, we need to impose $G(x) \geq 0$. Therefore, we will consider only the portion of the spacetime where $0 \le x \le x_1$. Importantly, we shall take the brane to be located at $x = 0$, which is always a totally umbilic surface in the C-metric. In the following we give a succinct description of these geometries and the braneworld construction of them, referring the reader to \cite{Emparan:1999wa,Emparan:1999fd,quBTZ} for more details.

The AdS$_4$ radius $\ell_4$ in \eqref{cmetric} is
\be \label{ell4}
    \frac{1}{\ell_4^2} = \frac{1}{\ell_3^2} + \frac{1}{\ell^2} \, ,
\ee
where $\ell_3$ is the AdS$_3$ radius on the brane and $\ell$ is directly related with the tension of the (two-sided)
brane as
\be \label{tau}
    \tau = \frac{1}{2\pi G_4 \ell} \, .
\ee
The other two parameters in the solution are dimensionless: $\kappa = \pm 1, 0$, which is a discrete one, and the real number $\mu$, which can be either a positive or a negative quantity. In the case $\kappa = -1$ and $\mu >0$ we recover the BTZ black hole on the brane, but we will carry out the study with arbitrary $\kappa$ and $\mu$ to include other interesting quantum black holes. We will eventually see from later results that we will get the holographic quantum corrections to the black hole.
Since our interest is the induced three-dimensional physics, we will typically keep $\ell_3$ fixed and then study the solutions for different values of $\mu$, $q$ and $\ell$. Then $\ell_4$ is a derived scale, as befits the notion that the bulk emerges from boundary physics.

The bulk is cut off at a brane at $x=0$. 
The Israel junction conditions for the geometry  determine the tension \eqref{tau} \cite{Israel:1966rt}. The junction conditions for the electromagnetic field can be easily obtained starting from Maxwell equations. We take $n^{\mu}$ to be a unit normal vector to the brane pointing towards increasing values of $x$ and $e_a^{\mu}$ is a basis for the tangent space to the brane. Therefore, projecting $F_{\mu \nu}$ on the brane as $F_{ab} \equiv F_{\mu \nu}e_a^{\mu}e_b^{\nu}$ and $f_a \equiv F_{\mu \nu}e_a^{\mu}n^{\nu}$, we get \cite{Lemos:2021jtm}
\begin{align}
   \left[ F_{ab} \right] & = F_{ab}^+ - F_{ab}^- =  0, \label{match_tangential} \\
   \left[ f_a \right] & = f_a^+ - f_a^- = 4\pi j_a \label{match_normal} ,
\end{align}
where the indices $a,b$ refer to the brane coordinates and the solution labelled as $+$ and $-$ refers to the spacetime at each side of the brane. The vector $j_a$ is the electromagnetic surface current.

The final regularity condition to consider is the smoothness of the geometry along the rotational symmetry axis $x = x_1$ where $G(x_1) = 0$. Absence of conical singularities requires that the coordinate $\phi$ is periodically identified as
\be \label{delta}
    \phi \sim \phi + 2\pi \Delta \, \quad \text{with} \quad \Delta = \dfrac{2}{\abs{G'(x_1)}} = \dfrac{2x_1}{\abs{-3+\kappa x_1^2 - q^2 x_1^4}} \, .
\ee

Let us now study the induced metric on the brane. We first ensure the coordinates are canonically normalized by rescaling
\be \label{coordinates_rescale}
    t = \Delta \bar{t}, \qquad  \phi = \Delta \bar{\phi}, \qquad r = \dfrac{\bar{r}}{\Delta} \, .
\ee
The metric on the brane is then
\be \label{metric_brane_canonic}
    ds^2|_{x=0}=  -H(\bar{r})d\bar{t}^2 + \dfrac{d\bar{r}^2}{H(\bar{r})} + r^2d\bar{\phi}^2
\ee
with
\be \label{Hr_brane}
    H(\bar{r}) = \dfrac{\bar{r}^2}{\ell_3^2} - 8\mathcal{G}_3M - \dfrac{\ell F(M,q)}{\bar{r}} + \dfrac{\ell^2Z(M,q)}{\bar{r}^2} \, .
\ee
We have identified
\begin{align} \label{FM}
    F(M,q) &= \mu \Delta^3 = 8 \dfrac{1-\kappa x_1^2-q^2x_1^4}{(3-\kappa x_1^2+q^2x_1^4)^3}\,  , \\ Z(M,q) &= q^2 \Delta^4 = q^2 \dfrac{16 x_1^4}{(-3+\kappa x_1^2 - q^2 x_1^4)^4} \, , 
\end{align}
and $\mathcal{G}_3$ is a `renormalized' Newton's constant, which is given by 
\be \label{G3_ren}
    \mathcal{G}_3 = \dfrac{\ell_4}{\ell}G_3 = \dfrac{1}{2\ell}G_4 \, .
\ee
This renormalization accounts for the effects that higher curvature terms induced on the brane have on the definition of the three-dimensional mass $M$. 
The identification proceeds in exactly the same way as in \cite{quBTZ}, and
\be \label{M_brane}
    M =  -\dfrac{\kappa}{8\mathcal{G}_3}\Delta^2 \, .
\ee
The charge, however, is more subtle and we will discuss it at greater length after we introduce the effective theory on the brane. 

The projected components of the electromagnetic tensor on the brane are 
\be \label{M_F}
    F_{\bar{r}\bar{t}} = \dfrac{2e \ell}{\gl \bar{r}^2}\Delta^2 \quad \text{and} \quad f_{\bar{\phi}}= -\dfrac{2 g \ell }{ \gl \bar{r}} \Delta^2 \, ,
\ee
which together with the matching condition (\ref{match_tangential}) allows us to determine that the induced current density on the brane takes the form 
\be \label{jphi_brane}
   j^{\bar{\phi}} =\dfrac{ g \ell  \Delta^2 }{\pi \gl \bar{r}^3}  \, .
\ee

Anticipating the more detailed analysis of backreaction below, the interpretation of these solutions as three-dimensional quantum backreacted black holes requires that $\ell$ be small---otherwise the three-dimensional effective theory breaks down. If one performs a perturbative expansion in $\ell$, then it follows from \eqref{cmetric} that the effects of the charge only enter at $\order{\ell^2}$. Therefore, any finite effects of the introduction of charge, such as we study next, require to consider finite values of $\ell$.

\subsection{Branches of Charged Quantum Black Holes}

The parameter $q$ in quantum black holes allows for \textit{extremal static black holes} on the brane, with degenerate, zero temperature horizons. These are a qualitatively new consequence of the introduction of charge, and understanding their main features allows to discern the behavior of the intermediate regimes of non-extremal charged quantum black holes. Therefore we will begin by examining the extremal limits of our black hole solutions.

Reaching extremality requires that the effects of charge backreaction on the horizon of the black hole are of the same size as those of chargeless origin. Since, as we have argued, charge backreaction appears at quadratic order, when the backreaction is non-zero but small the extremal horizon must also be small. There are two different ways in which this is possible:
\begin{itemize}
    \item $\kappa=+1$: the horizon appears when the backreaction dresses a horizonless defect of the kind in Sec.~\ref{condefs}. 
    \item  $\kappa=-1$: these solutions go over to massless BTZ black holes (whose area is zero) when backreaction is absent, as in Sec.~\ref{BTZdefs}. 
\end{itemize}

The distinction between $\kappa=\pm 1$ will be relevant in the discussions that follow.

\subsubsection{Extremal Quantum Black Holes}

Extremality requires a double root $r_+>0$ of $H(r)$, where $H(r_+)=H'(r_+)=0$. For this to be possible with $\ell_3$ and $\ell$ fixed, the parameters $\mu$ and $q$ must be related. It is actually convenient to use $r_+$ as a parameter, in terms of which the relation between $\mu$ and $q$ required for extremality is
\begin{align}\label{muqr0}
    \mu_0=\frac{2r_+}{\ell}\left(\kappa+\frac{2r_+^2}{\ell_3^2}\right)\,,\qquad
    q_0=\frac{r_+}{\ell}\sqrt{\kappa+\frac{3r_+^2}{\ell_3^2}}\,.
\end{align}
From the viewpoint of the brane it is clear that a non-zero quantum backreaction, $\ell\neq 0$, is needed in order to have an extremal horizon $r_+\neq 0$. The differences between $\kappa=\pm 1$ become apparent now:

\begin{itemize}
    \item When $\kappa=-1$ an extremal limit is possible in the absence of charge, $q_0=0$, with $r_+^2/\ell_3^2=1/3$. These are extremal hyperbolic black holes in the bulk with $\mu_0<0$. The extremal charged solutions all have $r_+^2/\ell_3^2> 1/3$. For $q_0\neq 0$, in the zero backreaction limit $\ell\to 0$ these black holes approach a massless BTZ black hole with a charged defect inside.\footnote{The limit is singular and subtle and will be clarified in \eqref{rplim}.} Nevertheless, when $\ell$ is small but finite the distortion away from the BTZ geometry must be large enough to render the horizon extremal; the angular distortion of the bulk horizon is large since in $G(x)$ the charge is not accompanied by factors of $\ell$.

    \item When $\kappa=+1$ and $\ell\neq 0$ the extremal horizon appears only when charge and its backreaction are included. When $\ell$ is small the black holes have small non-zero size $r_+\sim \mu_0 \ell/2 \sim q_0/\ell$ and from the bulk viewpoint they are well approximated as four-dimensional extremal Reissner-Nordström black holes---indeed, in the limit $r_+/\ell_3\to 0$ we recover the relation $q_0=\mu_0/2$ characteristic of them. 
    
\end{itemize}

In appendix~\ref{app:extremal} we give a more complete analysis of the different branches of solutions for the different roots $r_+$.

\paragraph{AdS$_2$ throat.} We now show that the geometry near the extremal horizon possesses an AdS$_2$ throat. We will actually include the leading order away from extremality, which is also captured in the throat limit. For this purpose, we keep $q=q_0$ fixed, take
\begin{align}\label{extremal_expansion}
    \mu=\mu_0\left(1+\frac{\varepsilon^2}{2}\right)\,,\qquad
    r=r_+\left(1+\varepsilon\rho\right)
\end{align}
and expand to leading order for $\varepsilon\ll 1$. Then 
\begin{align}\label{throat}
    -H(r)dt^2+\frac{dr^2}{H(r)}\quad\to\quad 
    -\varepsilon^2 r_+^2\frac{\rho^2-\rho_0^2}{L_2^2}dt^2+L_2^2\frac{d\rho^2}{\rho^2-\rho_0^2}
\end{align}
with
\begin{align}\label{ads2params}
    \rho_0^2=\frac{2r_+^2+ \kappa \ell_3^2}{6 r_+^2+\kappa \ell_3^2}\,,\qquad 
    L_2^2=\frac{r_+^2 \ell_3^2}{6 r_+^2+ \kappa \ell_3^2}\,.
\end{align}
We recognize in \eqref{throat} the geometry of thermal AdS$_2$ with radius $L_2$. In the parametrization we have chosen, the backreaction parameter $\ell$ is absent and indeed the expressions in \eqref{ads2params} are essentially the same as for the extremal RN-AdS$_4$ black holes \cite{Nayak:2018qej,Iliesiu:2020qvm}, which correspond to the limit $\ell\to\infty$ of our metrics.

In contrast to the RN-AdS solutions, the angular $S^2$ is not homogeneous. The function $G(x)$ for this part of the geometry can be expressed in the extremal solutions as 
\begin{align}
    G(x)=1-x^2\left(1+\frac{r_+}{\ell}x\right)^2-x^3\left( 4+3\frac{r_+}{\ell}x\right)\frac{r_+^3}{\ell \ell_3^2}\,,
\end{align}
which deviates from the form appropriate for a round sphere, $1-x^2$, since the acceleration of the black hole deforms the shape of its horizon. Furthermore, the AdS$_2$ factor is distorted in the angular direction $x$ due to the conformal factor $\propto 1/(\ell +x r_+)^2$ in \eqref{cmetric}. This feature is also present in the extremal rotating qBTZ solutions \cite{Kolanowski:2023hvh}.

\paragraph{Parameter space.} Later, when we study the thermodynamics of the general solutions, we will comment further on the entropy of the near-extremal black holes. We now characterize over which parameter regimes these solutions exist. The absence of $\ell$-dependence in \eqref{ads2params} suggests that the main qualitative features of the families of extremal solutions should remain the same for any finite nonzero value of the backreaction $\ell$. We will verify this in the following.

\begin{figure}[t]
  \centering
  \begin{minipage}[t]{0.45\textwidth}
    \includegraphics[width=\textwidth]{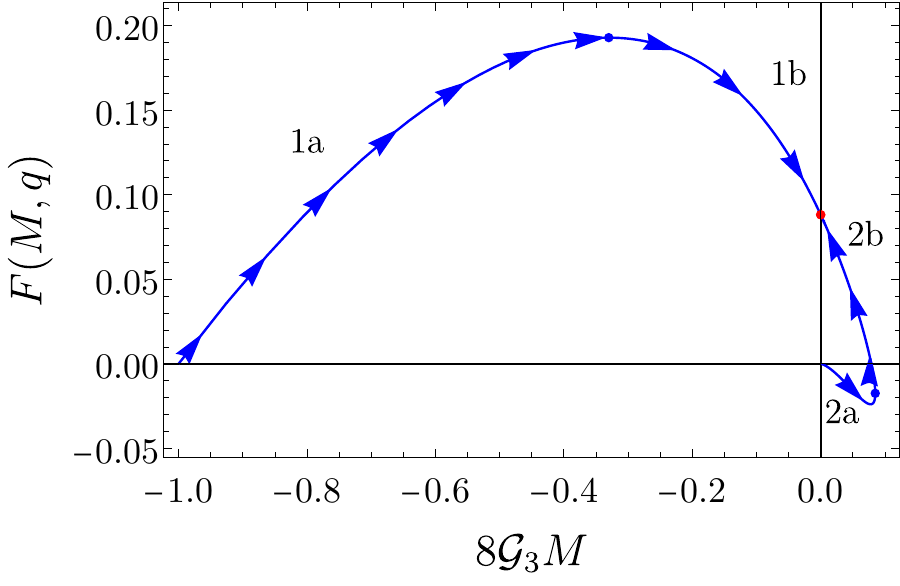}
  \end{minipage}
  \hfill
  \begin{minipage}[t]{0.45\textwidth}
    \includegraphics[width=\textwidth]{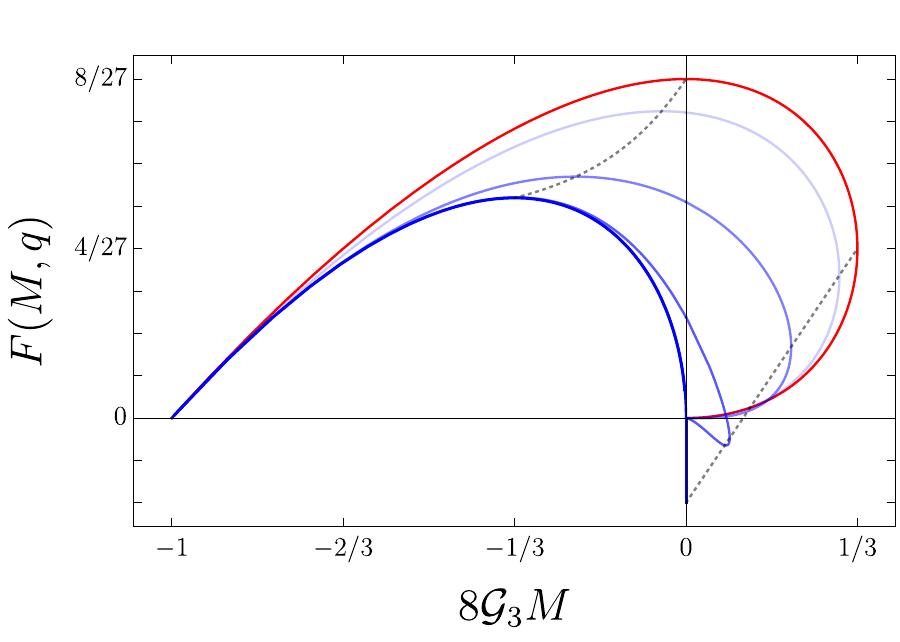}
  \end{minipage}
  \caption{\small Left: The function $F(M,q)$, which characterizes the holographic stress-energy tensor, for the four branches of an extremal charged quantum black hole for $\nu=1/5$. The curve is a parametric function of $q$ and the arrows indicate the direction of increasing $q$. Right: Branches of extremal quantum black holes for different values of $\nu$. The dark blue curve corresponds to the limit $\nu \to 0$ (which does not describe black holes but charged point defects), while the red curve corresponds to $\nu \to \infty$. The curves in between correspond to $\nu = 1/5, 3, 25$ in order of decreasing opacity. The dashed black lines show the points of maximum $F(M,q)$ (upper, center) and maximum $M$ (lower, right) as a function of $\nu$.}
  \label{fig:FM_extremal}
\end{figure}

 It is convenient to characterize these solutions as a function of the charge parameter $q$ and the dimensionless backreaction parameter introduced in \cite{quBTZ},
\be 
\nu = \frac{\ell}{\ell_3}\,.
\ee
In Figure~\ref{fig:FM_extremal} we show for a fixed value of $\nu$ a plot of $F(M, q)$ vs.~$M$ for the extremal quantum black holes. We break the curve into four segments which we label as 1a, 1b, 2a and 2b with the meeting points of each branch indicated by blue and red dots. Branches $1a$ and $1b$ both have $\kappa = +1$ while branches 2a and 2b have $\kappa = -1$.  The blue dots mark the transition between the different branches of solutions. The uppermost blue dot, which marks the meeting point of 1a and 1b, occurs at the maximum value of $F(M,q)$ for the extremal quantum black holes. The rightmost blue dot, which marks the meeting point of 2a and 2b, occurs at the maximum value of $M$. On the other hand, the red dot corresponds to the limit $q \to \infty$. Note that while the connection between branch 1b and 2b is smooth in the $(M, F)$ space, it is reached in the limit $q\to \infty$ independently from each side.

While we have fixed $\nu = 1/5$ in the left plot of Figure~\ref{fig:FM_extremal}, the qualitative shape of the curve is the same for any finite nonzero value of $\nu$. This is clear from Figure~\ref{fig:FM_extremal}, where we show the extremal quantum black holes for different values of $\nu$. All branches of extremal quantum black holes fall between two limiting cases: that given by the dark blue curve, which corresponds to the limit $\nu \to 0$, and the dark red curve, which corresponds to $\nu \to \infty$. Notably, the red curve perfectly agrees with the parameter ranges of the neutral quBTZ black hole.  In addition to these limiting cases, in the plot we include the branches of extremal quantum black holes for three finite, nonzero values of $\nu = 1/5, 2$ and $25$. The dashed black lines mark, as a function of $\nu$ the points where the different branches meet. That is, the points where $F(M,q)$ and $M$ reach their maxima for each value of $\nu$.

One notable feature of the extremal quantum black holes is the appearance of solutions with $F(M,q) < 0$, as we saw in Figure~\ref{fig:FM_extremal}. This occurs for the $\kappa = -1$ case when the value of $\mu$ at extremality is negative. Note that the ability to have $\mu < 0$ is a feature that is possible for the charged C-metric, but not the uncharged one. In the case where $q = 0$ and $\kappa = -1$, the existence of a horizon in the bulk requires $\mu < 0$. But this requirement is inconsistent with $G(x)$ having positive, finite roots. Hence the negative $\mu$ solutions cannot be consistently included in the ordinary quBTZ analysis. However, no such problem occurs when $q \neq 0$, allowing for the exploration of this parameter space. The same fact will extend to the non-extremal solutions, as we will see below. 

We will see below that the function $F(M,q)$ controls the stress-energy tensor of the CFT. In the absence of charge, the fact that $F (M)>0$  means that the energy density is negative and this is associated with the fact that the stress-energy tensor of the CFT is of Casimir type. The regimes where $F(M,q) < 0$ must then be regarded as states where the charge results in a positive energy density of the CFT  large enough to overcome the negative Casimir energy.

A simple analysis of the $\mu_{\rm ext}$---see Appendix~\ref{App:global_aspects}---shows that the extremal solutions have negative $\mu$ when 
\be 
    q < \frac{1}{\sqrt{12} \, \nu} \, .
\ee
Hence quantum black holes with negative $F(M,q)$ are present for all values of $\nu$ for some appropriate range of $q$. As it can be seen in Figure~\ref{fig:FM_extremal}, these black holes are all close to the $M=0$ BTZ black hole. When $\nu \to 0$, this portion of the curve becomes sharper, ultimately limiting to the portion of the dark blue curve that lies along the vertical axis. This is a non-perturbative feature, corresponding to the simultaneous limit $\nu \to 0$ and $q \to \infty$, and should not be regarded as physical solutions. On the other hand, as $\nu$ becomes large, the range of $q$ values for which $F(M,q)$ is negative decreases. This portion of the curve is still there, but becomes hardly visible for larger values of $\nu$.

\subsubsection{Non-extremal Quantum Black Holes}

\begin{figure}[t]
\centering
\includegraphics[width=0.75\textwidth]{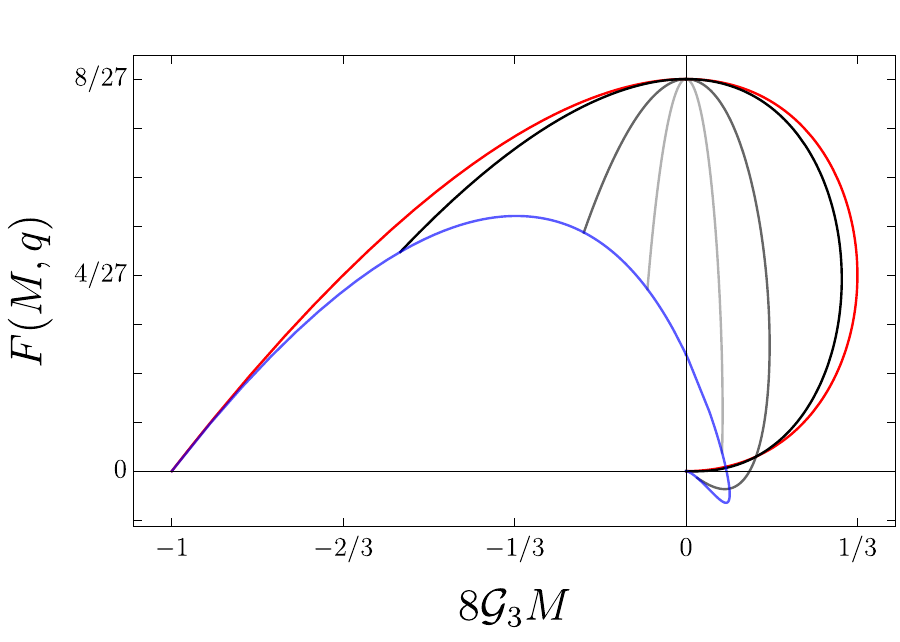}
\caption{\small The branches of non-extremal quantum black holes for $\nu = 1/5$. The red curve is the quBTZ black hole with $q = 0$, while the grey curves correspond to $q = 1/5, 1, 3$ in order of decreasing opacity. The blue curve traces out the extremal quantum black holes. Solutions with $M<0$ can be roughly viewed as the result of quantum backreaction on the horizonless charged defects of Sec.~\ref{condefs}. In the bulk they are approximately Reissner-Nordström black holes. Solutions with $M>0$ with large $F(M,Q)$ are the continuation of the previous ones,  but the branches with smaller $F(M,Q)$ are better viewed as the deformations by quantum backreaction of BTZ black holes of relatively small $M$. When the charge is large enough, its backreaction can result in $F(M,Q)<0$.}
\label{fig:FM_nonextremal}
\end{figure}

Having understood the structure of the families of extremal black holes, it is now simple to understand the non-extremal ones. We show a representative case in Figure~\ref{fig:FM_nonextremal} for $\nu = 1/5$. In this figure, the blue curve corresponds to the extremal black holes for the fixed value of $\nu$, while the red curve corresponds to the parameters traced out by the ordinary, $q=0$ quBTZ black hole. The three dark curves are black holes with fixed values of the charge parameter $q$. Roughly speaking, the curves traced out by the charged quantum black holes ``fill in'' the space between the quBTZ and extremal black hole curves. 

The branches of non-extremal quantum black holes share qualitative features with the neutral quBTZ black hole. First of all, the range of masses is always bounded---see appendix~\ref{App:global_aspects}---and a proper subset of the allowed masses for the quBTZ black hole. The minimum and maximum allowed masses are a function of both $q$ and $\nu$. The minimal mass is achieved for the $\kappa = +1$ black holes, and coincides with the extremal limit in all cases except for $q = 0$, in which case the minimum mass corresponds to global AdS$_3$. The maximum mass occurs for the $\kappa = -1$ black holes, but whether it is achieved by an extremal or non-extremal black hole depends on the particular values of $q$ and $\nu$. 

\subsubsection{The Effect of Backreaction}

We have so far discussed the properties of the quantum black holes localized on the brane. However, when $\nu \to 0$, the brane approaches the asymptotic AdS boundary. In this limit, the gravity and gauge field on the brane become non-dynamical and the situation reduces precisely to the one considered in section~\ref{sec:DoubleWicks}, with charged defects on a nondynamical boundary geometry. The solutions just discussed can be understood, in the context of braneworld holography, as the result of incorporating semiclassical backreaction effects to the configurations discussed in section~\ref{sec:DoubleWicks}.  

In the absence of backreaction, depending on the parameter values, the defects can reside either in a black hole or in a conical AdS$_3$ or Minkowski geometry.

For the configurations with $\kappa=-1$ that, without backreaction, correspond to a charged defect inside of a black hole, then backreaction always produces a charged quantum black hole on the brane. This is to be expected: the BTZ geometry at the boundary in the limit $\nu\to 0$ continues to possess a regular horizon after quantum backreaction is included. 

\begin{figure}
    \centering
    \includegraphics[width=0.65\textwidth]{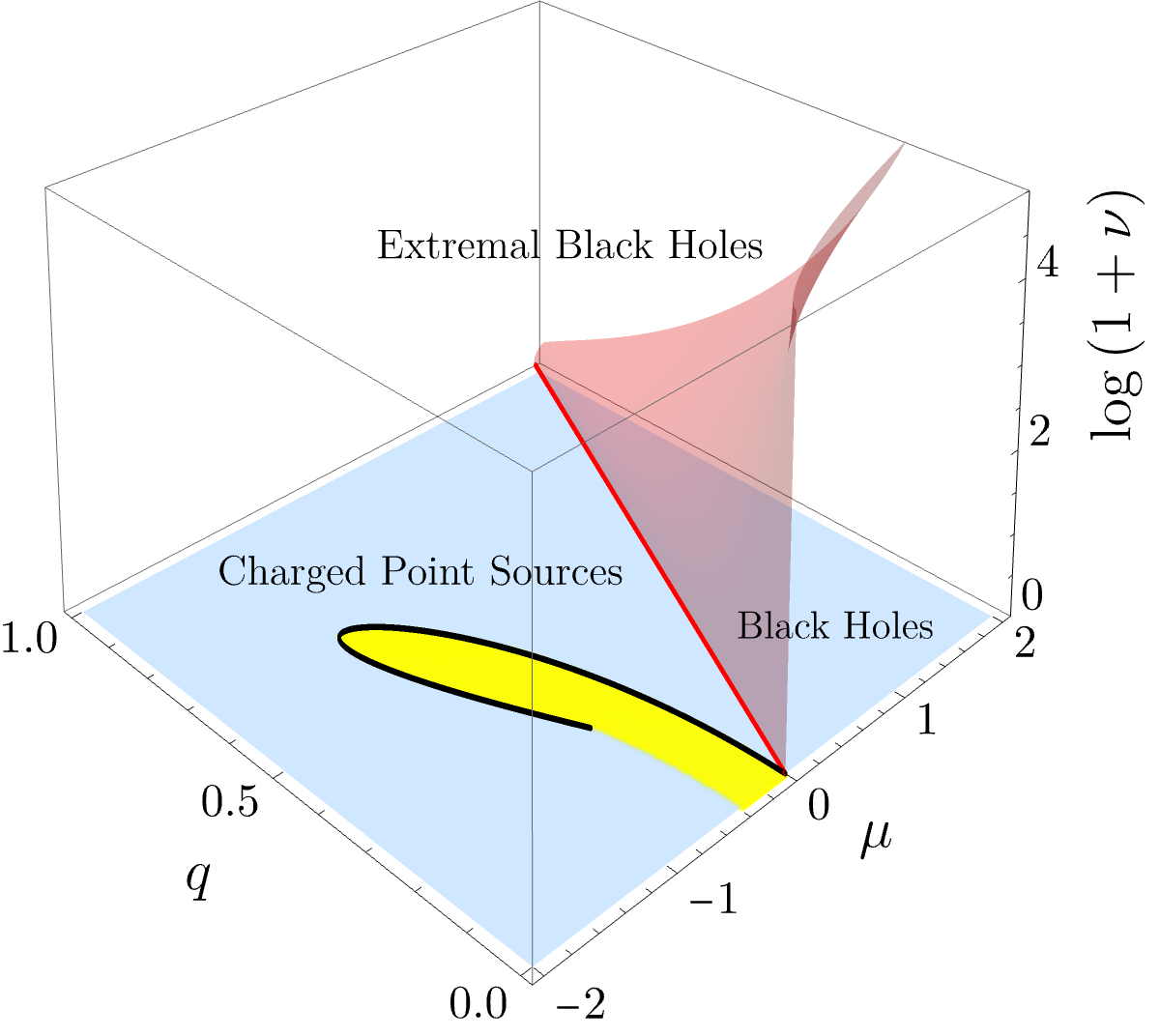}
    \caption{\small We reproduce Figure~\ref{defect_phase_space} adding the consequences of the backreaction on the solutions with $\kappa=+1$. The backreaction parameter $\nu$ increases along the vertical axis. In the first section (corresponding to the no-backreaction plane $\nu=0$), the blue-shaded region were conical defects, the yellow-shaded region were conical excesses, and the black line were smooth boundary geometries. Their fate when backreaction is accounted for depends on which side of the red surface they lie on. This surface, which intersects the plane $\nu=0$ along the red line $q=\mu/2$, indicates extremal black holes that separate between naked timelike singularities, i.e., charged point sources (above the extremal surface), and charged quantum black holes (below the extremal surface). When a $\nu=0$ solution above the red line, with $q\geq \mu/2$, is backreacted, it always results in a charged point source. In particular, the smooth geometries of the first section, on the black line, all turn into singular charged point sources. Solutions below the red line, with $q<\mu/2$, produce charged black holes if the backreaction is small enough (smaller for larger charge) and charged point sources if it is larger.}
    \label{fig:backreaction}
\end{figure}

Figure~\ref{fig:backreaction} shows what happens when backreaction is implemented in the more subtle situation with $\kappa=+1$, when the $\nu = 0$ geometries are charged defects in AdS$_3$ or Minkowski. In this case the horizon can only appear due to quantum backreaction, but (in contrast to the neutral case) it need not do so if the charge is large enough. That is, the effect of backreaction depends on the ratio of the charge and mass. We highlight this dependence in Figure~\ref{fig:backreaction}. If the non-backreacted system has $q \geq \mu/2$, then backreaction will always result in a naked timelike singularity.  If $q < \mu/2$, then the backreaction will produce a quantum black hole if the backreaction is not too large, but as the backreaction grows it reaches a value (which is smaller for larger $q$) at which an extremal black hole eventually forms. Stronger backreaction results in naked singularities. 

The solutions with a naked singularity are not necessarily unphysical. The regime of $q>\mu/2$ is analogous (when not actually the same) as the regime of $Q>M$ in the Reissner-Nordström solutions. These are interpreted as correctly describing the gravitational field of a charged particle at least down to the distances where quantum electrodynamics (not quantum gravity), through virtual production of charged pairs, modifies the stress tensor of the electromagnetic field making it less singular. Our solutions must be interpreted in the same manner. Observe that we find that backreaction creates a naked singularity even in the situations with $\alpha=1$ where, in the absence of backreaction, the geometry where the defect lives is smooth. The reason is that the electric potential $\propto a/r$ is singular at $r=0$, and thus its stress tensor diverges and creates a curvature singularity. Like in the Reissner-Nordström case, these solutions must be regarded as charged pointlike sources.

\subsection{Effective Brane Theory and CFT Response}

The three-dimensional black holes we have been discussing are to be understood as solutions of an effective theory obtained by integrating out the degrees of freedom above a cutoff energy $1/\ell$,
\begin{align}
    I = \dfrac{1}{16\pi G_4} &\bigg[ \int_{\rho \geq \ell} d^{4}x \sqrt{-g} (R-2\Lambda) + \int_{\rho=\ell} d^3x \sqrt{-h}K - 16\pi G_4 \int_{\rho=\ell} d^3x \sqrt{-h}\tau 
    \nonumber\\
    &- \dfrac{\gl^2}{4} \int_{\rho\geq \ell}d^{4}x \sqrt{-g}F_{\mu\nu}F^{\mu\nu}+ 16 \pi G_4 \int_{\rho=\ell} d^3x \sqrt{-h} A_{a}j^{a} \bigg] \, ,
    \label{action}
\end{align}
where $\rho \rightarrow 0$ is the boundary of AdS and the brane is located near the boundary, at $\rho = \ell$. The above is to be evaluated in a small $\ell$ expansion.  The resulting action is
\beq \label{effective_action}
    I = \dfrac{\ell_4}{8\pi G_4}\int d^3x \sqrt{-h} \left[ \dfrac{4}{\ell_4^2}\left( 1-\dfrac{\ell_4}{\ell} \right) + R + \ell_4^2\left( \dfrac{3}{8}R^2 - R_{ab}R^{ab} \right) + \mathcal{O}(\ell_4^3 ) \right] + I_{\text{EM}}+ I_{\text{CFT}} \, .
\eeq
In the above, the first terms are the ones that would be cancelled by the counterterms of holographic renormalization of AdS Einstein gravity (with an additional factor of 2 due to the $\mathbb{Z}_2$ symmetry of the configuration). Additionally, we have $I_{\rm CFT}$, which is the action of the conformal fields residing on the brane and, in our setup, are defined holographically by the four-dimensional bulk. Finally, due to the presence of the Maxwell term in the four-dimensional action, there arise also $U(1)$ fields on the brane,
\beq
    I_{\text{EM}} = 2 \int d^3x \sqrt{-h} A_aj^a + I^{\text{ct}}_{\text{EM}}
\eeq
where $I^{\text{ct}}_{\text{EM}}$ are the counterterms of the electromagnetic action. Since there are no divergences associated to the Maxwell term in four-dimensional space-time, these counterterms are not required in the usual prescription of holographic renormalization. However, since here the brane is at a finite distance from the boundary, these terms make essential contributions to the effective field theory on the brane. These terms were thoroughly studied in~\cite{Taylor:2000xw}, leading to the result
\begin{align} \label{maxwell_counterterms}
    I^{\text{ct}}_{\text{EM}} =  \dfrac{\ell_4 \gl^2}{8\pi G_4}  \int & d^3x \sqrt{-h} \bigg[ - \dfrac{5}{16}F^2 + \ell_4^2 \bigg(\dfrac{1}{288}RF^2 - \dfrac{5}{8} R^a_b F_{ac}F^{bc}  \nonumber \\ 
     & + \dfrac{3}{98}F^{ab}\left( \nabla_b\nabla^c F_{ca} - \nabla_a\nabla^c F_{cb} \right) + \dfrac{5}{24}  \nabla_aF^{ab}\nabla_c{F^c}_b \bigg) + \mathcal{O}(\ell_4^3 )\bigg] \, .
\end{align}

Strictly speaking, the brane theory should be treated in the sense of effective field theory~\cite{Omiya:2021olc, Karch:2022rvr, Bueno:2022log, Neuenfeld:2023svs}, with the relevant length scale set by the parameter $\ell$. Considering this to be a small parameter, i.e. $\ell \ll \ell_3$, we can expand $\ell_4 \sim \ell$ so that the leading terms in the action read
\begin{align} \label{complete_effective_action}
    I = &\dfrac{1}{16\pi G_3}\int d^3x \sqrt{-h} \Bigg\{ \dfrac{2}{L_3^2} + R - \frac{\glbrane^2}{4} F^2 + 16 \pi G_3 A_aj^a  + \ell^2 \left( \dfrac{3}{8}R^2 - R_{ab}R^{ab}  \right)  + \mathcal{O}(\ell^4 )\Bigg\} \nonumber \\
    &+ I_{\rm CFT} \, ,
\end{align}
where we have also identified the bare three-dimensional Newton's constant, three-dimensional cosmological constant, and gauge coupling as
\beq
    G_3=\dfrac{1}{2\ell_4}G_4 \, , \qquad  \dfrac{1}{L_3^2}=\dfrac{2}{\ell_4^2}\left( 1-\dfrac{\ell_4}{\ell} \right) = \dfrac{1}{\ell_3^2} \left( 1 + \dfrac{\ell^2}{4\ell_3^2} + \mathcal{O}(\ell^4) \right) \, , \qquad  \glbrane^2 = \frac{16 \pi G_3}{g_3^2} \, .
\eeq
Note that we have absorbed the numerical factors above into the definition of $G_3$ and the gauge field length scale so that the action takes the conventional form. The relationship between the central charge $c$ of the CFT and the parameter $\ell$  of the three-dimensional theory is unaltered and given by \cite{quBTZ}
\beq \label{c}
    \ell = 2cG_3 \left( 1 + \mathcal{O}(cG_3/L_3)^2 \right) \, .
\eeq

The three-dimensional gauge coupling is related to the four-dimensional one by 
\be\label{g3def} 
g_3^2 =  \frac{2}{5} \frac{g_\star^2}{\ell_4} \approx \frac{2}{5} \frac{g_\star^2}{\ell}   \, .
\ee
The numerical factors ensure that the Maxwell sector of the theory has the conventional normalization,
\be 
I_{\rm EM} = \int d^3 x \sqrt{-h} \left[-\frac{1}{4 g_3^2} F^2 + A_aj^a + \cdots  \right] \, ,
\ee
(where the dots indicate the various higher-derivative operators that appear in the effective action). The appearance of $\ell$ in eq.~\eqref{g3def} makes manifest that the gauge field becomes non-dynamical in the limit $\ell \to 0$, when the brane is pushed to the asymptotic AdS boundary.

The metric and gauge field~\eqref{metric_brane_canonic} solve  the --- Einstein-Maxwell equations that follow from the above effective theory.  We can identify the holographic CFT stress tensor as the right-hand side of the gravitational equations derived from the action (\ref{complete_effective_action}), which take the form
\begin{align}
    8\pi G_3 \langle  T_{ab}\rangle  = &R_{ab} - \dfrac{1}{2}h_{ab}\left( R+\dfrac{2}{L_3^2} \right)  - \frac{\glbrane^2}{2} \left(F_a^c F_{b c} - \frac{1}{4} h_{ab} F^2 \right) +  16 \pi G_3 A_c j^c h_{ab}   \nonumber \\
    +& \ell^2 \left[ 4{R_a}^cR_{bc} -\dfrac{9}{4}RR_{ab}-\nabla^2 R_{ab} + \dfrac{1}{4}\nabla_a\nabla_b R \right. \nonumber \\
    & \ \ \ \ \left. + h_{ab}\left( \dfrac{13}{8} R^2 - 3R_{cd}R^{cd} + \dfrac{1}{2}\nabla^2R \right) \right] + \mathcal{O}(\ell^3) \, .
\end{align}
Computing these terms for the exact metric (\ref{metric_brane_canonic}) coordinates and taking the limit of small backreaction $\ell\rightarrow 0$ with (\ref{c}) we get at leading order
\beq 
    \langle {T^a}_{b}\rangle  \sim \dfrac{cF(M)}{\bar{r}^3} \text{diag}\{ 1,1,-2 \} + \mathcal{O}(\ell^2) \, . 
\eeq
The effects of charge appear at $\order{\ell^2}$ and therefore to leading order the function $F(M,q)$ reduces to the same $F(M)$ as computed for neutral quantum black holes in \cite{Emparan:2020rnp}. For the extremal solutions the small-$\ell$ expansion of the stress-energy tensor is expected to break down, since the curvature and/or the gauge field become large near the horizon, e.g., ${R_a}^cR_{bc}\sim \ell^{-2}$ or $F_a^c F_{b c}\sim \glbrane^{-2}$. 

From the brane action we can also compute the current response, which sources the semiclassical Maxwell equations. The result is
\begin{align}
    \langle J^{b}\rangle  =& j^b + \dfrac{\glbrane^2}{16 \pi G_3}\left[\nabla_a F^{ba} \right. + \frac{16}{5} \ell^2 \left( 
    -\dfrac{1}{72}R\nabla_a F^{ba} \right. + \dfrac{11}{18}{F^b}_a\nabla^aR + \dfrac{209}{294}R^{ba} \nabla_c {F_a}^c   \nonumber \\
     &+\dfrac{5}{4}R^{ac}\nabla_c {F^b}_a + \dfrac{5}{4}F^{ac}\nabla_c {R^b}_a + \dfrac{317}{588} \nabla_a\nabla^a\nabla_c F^{bc} + \left. \left. \dfrac{317}{588} \nabla^b\nabla^c\nabla_a F^{ac} \right) + \mathcal{O}(\ell^3) \right] \, .
    \label{effective_current_density}
\end{align}
Computing the terms for the Maxwell tensor at the brane (\ref{M_F}) and (\ref{jphi_brane}), obtain the following components of the current density at leading order in $\ell$
\be  \label{J_eff} 
    \langle J^\phi\rangle = \frac{\ell}{\gl} \frac{g \Delta^2 }{\pi \bar{r}^3} \, ,\qquad 
    \langle J^t\rangle = -\dfrac{\ell \glbrane^2}{8 \pi G_3 \gl} \dfrac{e\Delta^2}{\bar{r}^3}\,.
\ee
We can re-express holographic current density in terms of CFT parameters giving
\be 
\langle J^\phi\rangle \propto \frac{g \gs \sqrt{c}}{\bar{r}^3} \, , \qquad \langle J^t\rangle \propto \frac{e \ell \sqrt{c}}{\gs \bar{r}^3}  \, .
\ee
We see that the azimuthal component is independent of the backreaction parameter $\ell$, while the temporal component has a linear $\ell$-dependence. As such, in the limit $\ell \to 0$ we recover the same behaviour as obtained in section~\ref{sec:DoubleWicks}.

Finally, let us comment on the notion of charge from the three-dimensional perspective. In three dimensions, an electric monopole charge arises from a potential with a logarithmic dependence on the radial variable, i.e.
\be 
A_{t}^{\rm monopole} \sim Q \log r \, .
\ee
Here, the gauge field does not possess any such logarithmic behaviour. This can be understood as a consequence of the fact that the four-dimensional gauge field does not localize on the brane in the same way that gravity does---see~\cite{Bajc:1999mh} for a related discussion in the context of Randall-Sundrum braneworlds. To compute the charge one should include the full, infinite tower of higher-derivative corrections to the Einstein-Maxwell theory, and study the conserved $U(1)$ current of the resulting theory.\footnote{See~\cite{Bueno:2023qpr} for an explanation of the infinite derivative theory of gravity arising in the braneworld construction.} As we will see below, based on thermodynamic reasoning, the result should be equivalent to the charge computed in the higher-dimensional spacetime.

\subsection{Thermodynamics}

\begin{figure}[t]
  \centering
  \begin{minipage}[t]{0.45\textwidth}
    \includegraphics[width=\textwidth]{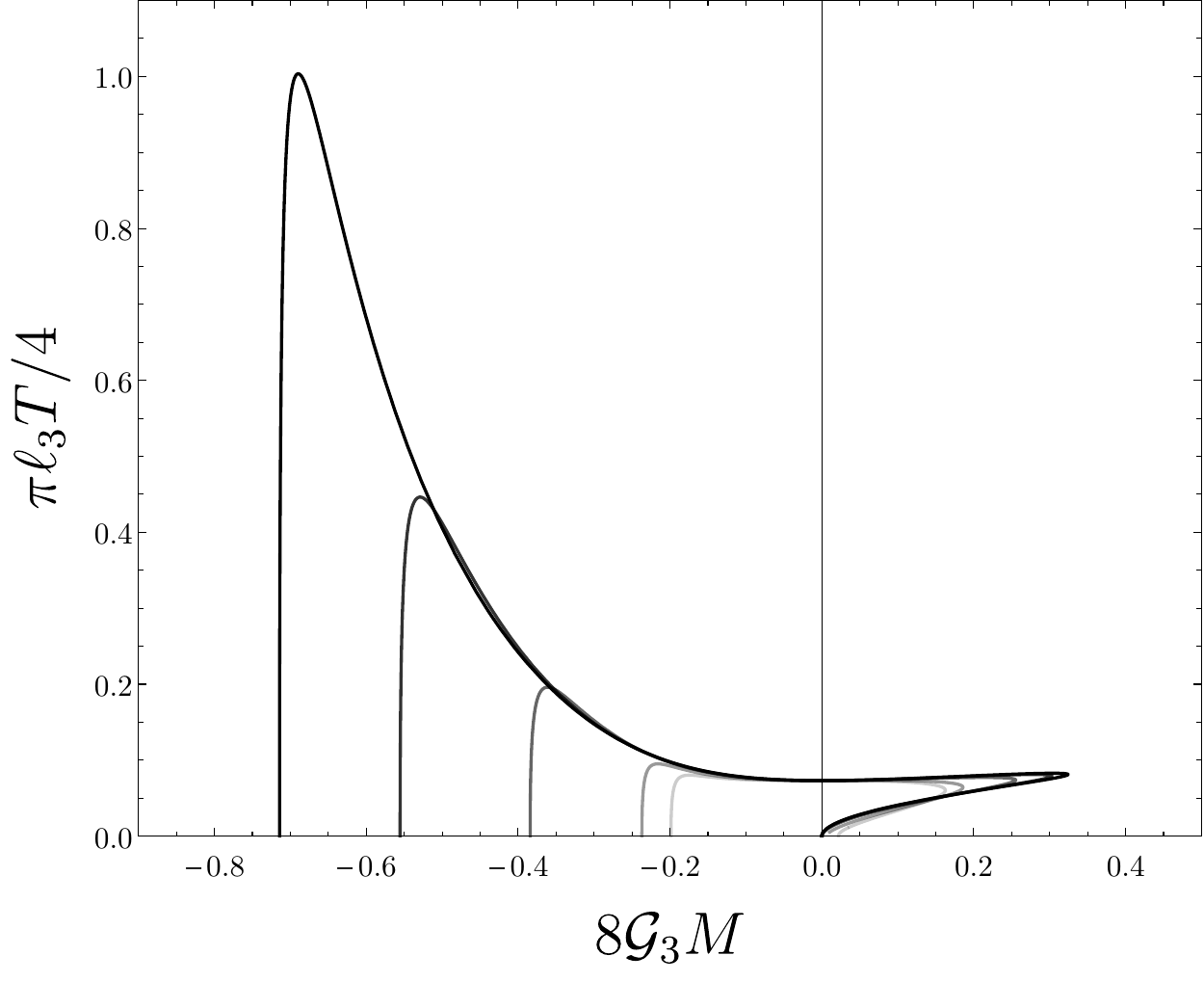}
  \end{minipage}
  \hfill
  \begin{minipage}[t]{0.45\textwidth}
    \includegraphics[width=\textwidth]{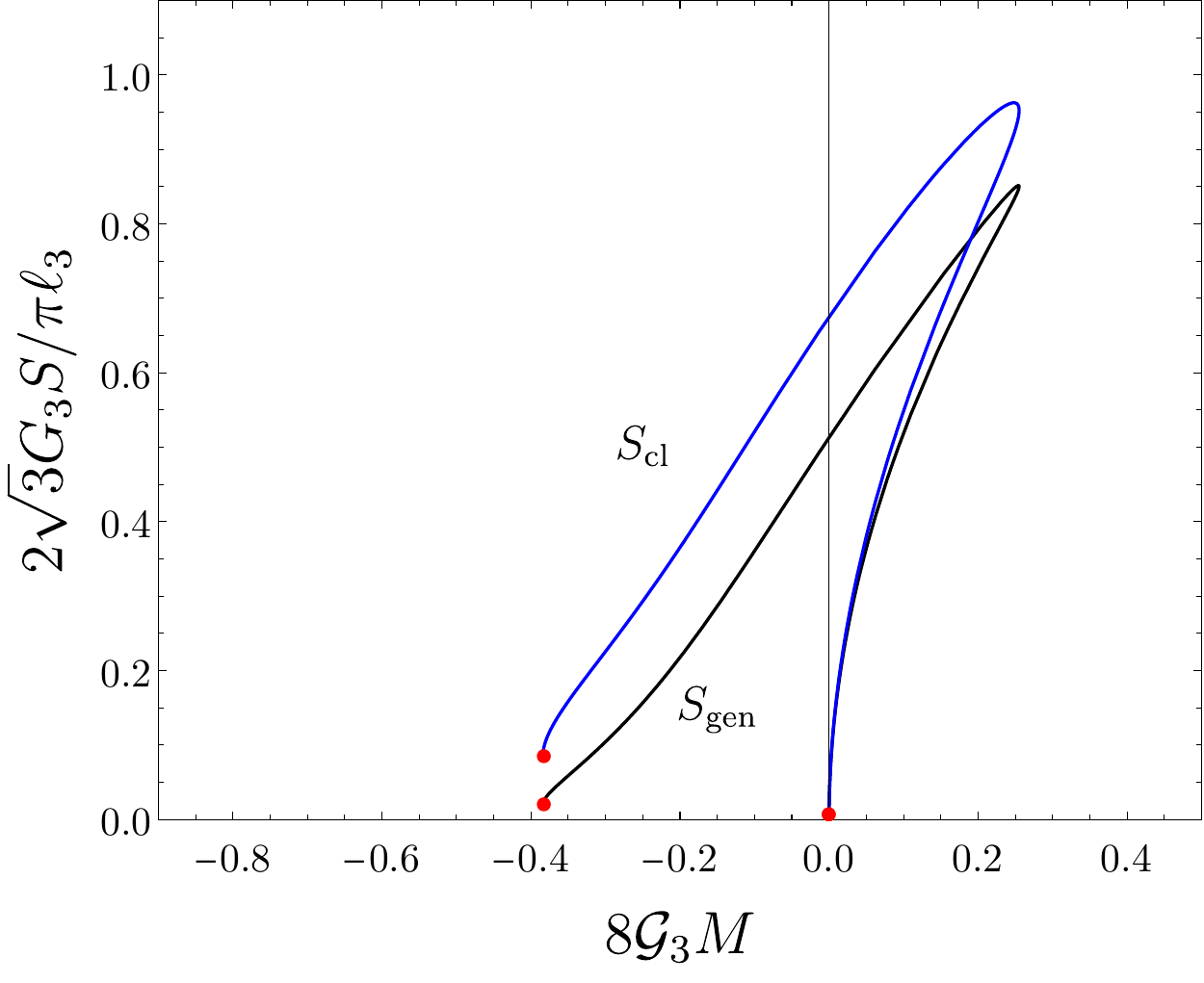}
  \end{minipage}
  \caption{\small Left: Temperature of the black hole as a function of the mass $M$ with $\nu=1/5$. Each curve represents a different value of $q=0.1,0.2,0.4,0.8,1$. The darker curves corresponds to smaller values of $q$. Right: Generalized entropy of a black hole as a function of the mass for $q=0.4$ and $\nu=1/5$. The blue line corresponds to the generalized entropy and the red one to the classical entropy. The red dots indicate the extremal black hole.}
  \label{fig:TandS}
\end{figure}

Finally, we discuss the thermodynamics of the quantum black holes. The determination of the mass, temperature, and entropy is straightforward as it formally follows from the general discussion of~\cite{quBTZ}. The black hole mass $M$ computed on the brane taking into account the higher-curvature corrections to the gravitational action, was already given in \eqref{M_brane}. The temperature and generalized entropy are 
\begin{align}
    T &= \frac{\Delta H'(r_+)}{4\pi} = \frac{3\Delta}{4 \pi x_1 \ell_3} \left[\frac{1}{z} + \frac{\kappa z x_1^2}{3}  - \frac{\nu^2 q^2 x_1^4 z^3}{3} \right] \, ,
    \\
    S_{\rm gen} &= \frac{2}{4 G_4} \int_0^{2 \pi \Delta } d\phi \int_0^{x_1}  \frac{\ell^2 r_+^2}{(\ell + x r_+)^2}  dx = \frac{\pi \ell_3}{2 G_3} \frac{\Delta  \sqrt{1+\nu^2}}{x_1 z \left(1+\nu z\right)}  \, ,
    \label{entropy}
\end{align}
where we have defined 
\be \label{x1_z}
    z = \dfrac{\ell_3}{r_+x_1} \, .
\ee
The temperature $T$ is computed relative to the timelike Killing vector, $\partial/\partial {\bar t}$, which has the correct normalization for a black hole on the brane. The entropy $S_{\rm gen}$ is identified with twice the entropy of the four-dimensional black hole, with the factor of two arising from the $\mathbb{Z}_2$ symmetry. From the brane perspective, this includes both the Wald entropy (including contributions from higher-derivative corrections) along with the entanglement entropy of the quantum fields.

We show the behaviour of the temperature and entropy in Figure~\ref{fig:TandS} for representative values of the parameters. In the case of the entropy, we also compare it with the `classical' Bekenstein-Hawking entropy, which is the area of the black hole horizon on the brane,
\be
    S_{\text{cl}} = \dfrac{1}{4G_3}2\pi r_+ \Delta = \dfrac{1+\nu z}{\sqrt{1+\nu^2}}S_{\text{gen}}  \, .
\ee
The behaviour of the temperature in the regime of positive masses is qualitatively similar to the uncharged quBTZ black hole. Differences arise for negative mass. The temperature of the uncharged quBTZ black hole grows monotonically as the mass is decreased toward its minimum value. Contrast this with the charged quantum black hole, for which the temperature reaches a maximum and the rapidly tends to zero in the extremal limit. The behaviour of the generalized entropy is qualitatively the same for both the uncharged and charged quBTZ metrics. The only major difference is that, in the charged case, the curve terminates at finite entropy when extremal solutions are reached.

As for the chemical potential, in the electric case we have chosen a gauge so that it vanishes on the black hole horizon. The electric chemical potential is the boundary value of the gauge field
\be
    \mu_e =  \frac{ 2\ell e \Delta}{ r_+ \gl} \, . 
\ee
The magnetic potential has exactly the same form upon changing $e \to g$. 

The most subtle aspect is the determination of the `charge' that enters into the first law. Ultimately, this is equivalent to the charge from the four-dimensional bulk. As discussed above, since the gauge field does not localize on the brane, a determination of the charge from the brane action requires a full resummation of the infinite tower of corrections appearing in the effective action. The four-dimensional bulk performs this resummation, with the result that
\be
    Q_e = \frac{2}{g_{\star}^2} \int \star F =   \dfrac{8 \pi \ell e\Delta x_1}{g_\star^2 \gl}  \, ,
\ee
where the prefactor of $2$ arises from the $\mathbb{Z}_2$ symmetry of the configuration. The magnetic charge has the same form with $e \to g$. 

The thermodynamics can be put into a form that is quite amenable to computation by introducing the notation
\be
\gamma = q x_1^2 \, , \qquad \gamma_e = e x_1^2 \, , \qquad \gamma_g = g x_1^2 \, .
\ee
The thermodynamic parameters then read
\begin{align}
    M &= \frac{\sqrt{1+\nu^{2}}\, \left(z \nu +1\right) \left(1+\gamma^{2} \nu^{2} z^{4}+\nu  \left(\gamma^{2}-1\right) z^{3}\right) z^{2}}{2 G_3 \left(1+\gamma^{2} \nu^{2} z^{4}+\left(2 \gamma^{2}+2\right) \nu  \,z^{3}+\left(\gamma^{2}+3\right) z^{2}\right)^{2}}
    \\
    T &= \frac{\left(2-\gamma^{2} \nu^{3} z^{5}-2 \gamma^{2} \nu^{2} z^{4}-\nu  \left(\gamma^{2}+1\right) z^{3}-3 z \nu \right) z}{2 \pi  \ell_3 \left(1+\gamma^{2} \nu^{2} z^{4}+\left(2 \gamma^{2}+2\right) \nu  \,z^{3}+\left(\gamma^{2}+3\right) z^{2}\right) } \, ,
    \\
    S_{\rm gen} &= \frac{\pi  \ell_3}{G_3} \frac{\sqrt{1+\nu^{2}}\, z}{\left(1+\gamma^{2} \nu^{2} z^{4}+\left(2 \gamma^{2} \nu +2 \nu \right) z^{3}+\left(\gamma^{2}+3\right) z^{2}\right)} \, ,
    \\
    \mu_e &= \sqrt{\frac{5 g_3^2}{4 \pi G_3}} \frac{\nu  \gamma_e  \,z^{3} \left(z \nu +1\right)}{\left(1+\gamma^{2} \nu^{2} z^{4}+\left(2 \gamma^{2}+2\right) \nu  \,z^{3}+\left( \gamma^{2}+3\right) z^{2}\right)} \, ,
    \\
    Q_e &=  \sqrt{\frac{16 \pi }{5 g_3^2 G_3}} \frac{\gamma_e  z^{2} \left(z \nu +1\right) \sqrt{1+\nu^{2}}}{ \left(1+\gamma^{2} \nu^{2} z^{4}+2\nu \left(\gamma^{2} + 1\right) z^{3}+\left(\gamma^{2}+3\right) z^{2}\right)} \, .
\end{align}
The magnetic chemical potential and charge are given simply by replacing $\gamma_e \to \gamma_g$ in the last two expressions above. Taken all together, these quantities satisfy the first law of thermodynamics
\be
    dM = TdS_{\rm gen} + \mu_e dQ_e + \mu_g dQ_g \,, 
\ee
where the variations are computed at fixed values of $G_3$, $g_3$, $\ell_3$ and $\nu$. While the parameters introduced above make the verification of the first law rather simple, care must be exercised in determining the allowed range of parameters which follow from the combined regularity constraints of the $H(r)$ and $G(x)$ polynomials. 

While it would be interesting to study in full detail the thermodynamic properties of the solutions---see~\cite{Frassino:2022zaz, Frassino:2023wpc, Johnson:2023dtf, HosseiniMansoori:2024bfi, Wu:2024txe} for recent progress on this in the case of the quBTZ black hole---, here we will focus our attention on the solutions near extremality. As mentioned in the introduction, at sufficiently low temperatures the physics of the near-extremal solutions is dominated by quantum gravitational fluctuations of the AdS$_2$ throat~\cite{Iliesiu:2020qvm}. Our analysis includes the quantum effects of the CFT, which dominate over the quantum fluctuations of the metric only away from the throat. Here we provide some commentary on when the semiclassical description of the throat can be expected to break down. 

\begin{figure}[t]
  \centering
  \begin{minipage}[t]{0.47\textwidth}
    \includegraphics[width=\textwidth]{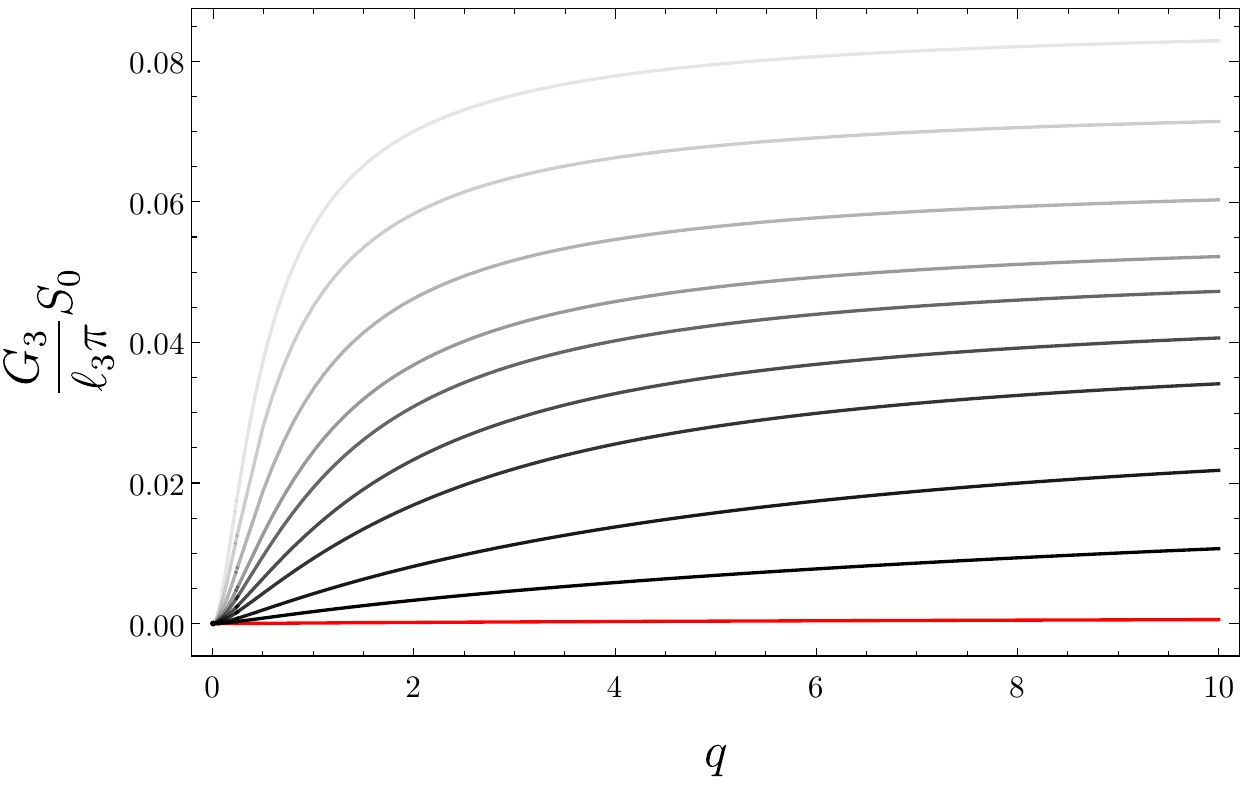}
  \end{minipage}
  \hfill
  \begin{minipage}[t]{0.47\textwidth}
    \includegraphics[width=\textwidth]{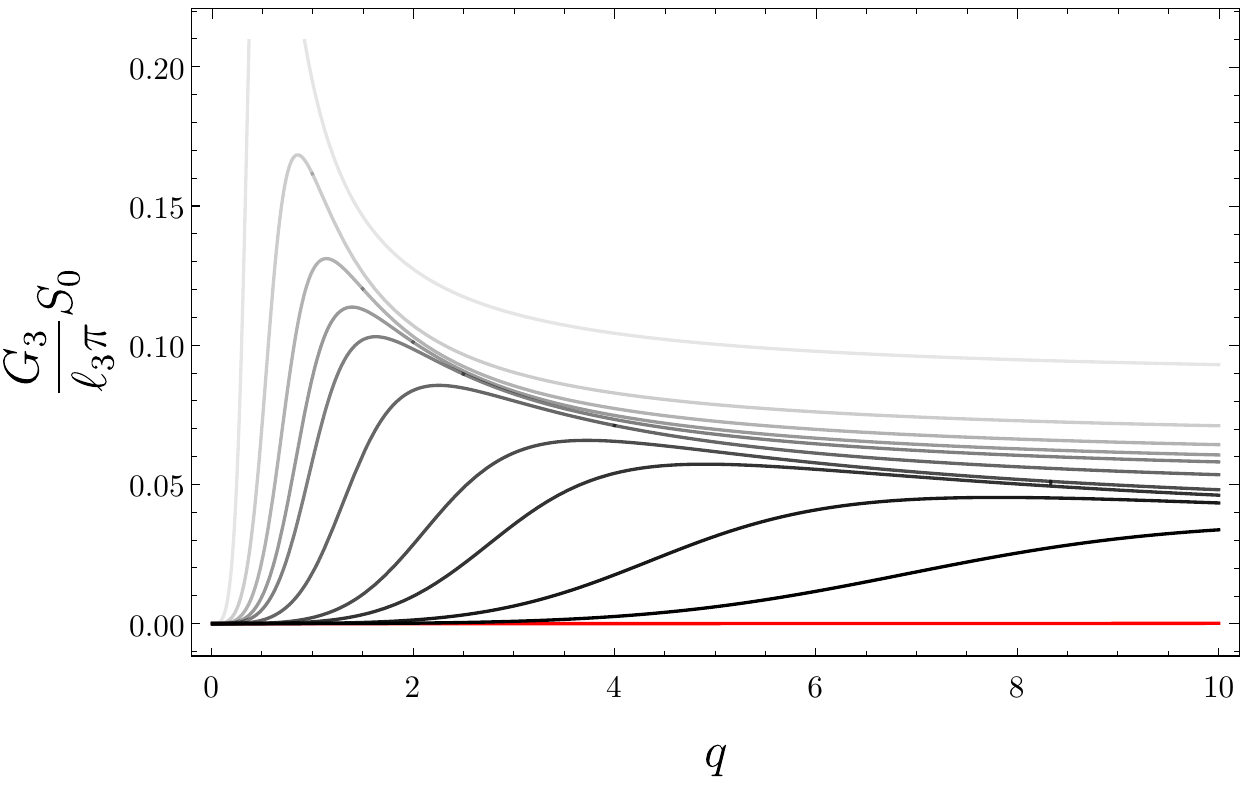}
  \end{minipage}
  \caption{\small Entropy for extremal black holes $S_0$ as a function of $q$ for $\kappa =+1$ (left) and $\kappa =-1$ (right). The former are dressed charged defects that in the bulk are Reissner-Nordström-like, and the latter are like small-mass-BTZ driven to extremality with charge. The curves with lighter colors correspond to higher values of the backreaction $\nu$. The red line corresponds to the zero backreaction limit $\nu \rightarrow 0$. }
  \label{fig:S0}
\end{figure}

The entropy of a near-extremal black hole can be expanded at low temperatures as 
\begin{align}\label{nextS}
    S = S_0 + \frac{4 \pi^2}{M_{\rm gap}} T + \dots \, ,
\end{align}
where $S_0$ is the saddle-point entropy of the extremal solution (given by the four-dimensional area) and $M_{\rm gap}$ is an energy scale dependent on the characteristics of the solution.  When the temperature is below the scale set by $M_{\rm gap}$ quantum gravitational fluctuations become important.  
Expanding the entropy and temperature near extremality allows us to identify 
\begin{align}
M_{\rm gap} &= \frac{4 G_3}{L_2^2} \frac{\left(1+\nu z \right)^2}{\sqrt{1+\nu^2} \left(1+2\nu z\right)} \, ,
\end{align}
where $z$ is evaluated for the extremal solution. We have expressed the result in terms of the curvature radius of the AdS$_2$ throat $L_2$---see eq.~\eqref{ads2params}---for simplicity. 

As we have explained, quantum fluctuations of the throat cancel the saddle-point value of the entropy $S_0$ and drive the extremal black hole entropy to zero. However, the value of $S_0$ still has interest as a measure of the size of the black hole in Planck units. In Figure~\ref{fig:S0} we plot it as a function of the charge $q$ for different values of the backreaction $\nu$ and $\kappa=\pm 1$. 
For context, recall that in the limit of vanishing backreaction, the quantum black holes with $\kappa = +1$ reduce to charged point particles at the boundary, while for $\kappa = -1$ they were BTZ black holes with a charged defect inside. Gravitational backreaction dresses the former with a horizon that, in the bulk, resembles that of a Reissner-Nordström black hole. The solutions with $\kappa=-1$ that are extremal must be regarded as small-mass BTZ black holes driven to extremality by the addition of quantum charge.

The left plot is for $\kappa=+1$, i.e., dressed charged defects that in the bulk are roughly like Reissner-Nordström black holes. These behave as expected: as the charge grows, and also as the backreaction increases, the size of the black hole, and hence $S_0$, always grows.
The right plot in Figure~\ref{fig:S0} is for $\kappa=-1$. Increasing the backreaction (for fixed $q$) always increases the black hole size $S_0$. This is expected, but the effect of adding charge is less obvious: larger $q$ makes the horizon size grow, as might be expected, but only up to value where $S_0$ reaches a maximum and then slowly decreases. 

\begin{figure}[t]
  \centering
  \begin{minipage}[t]{0.45\textwidth}
    \includegraphics[width=\textwidth]{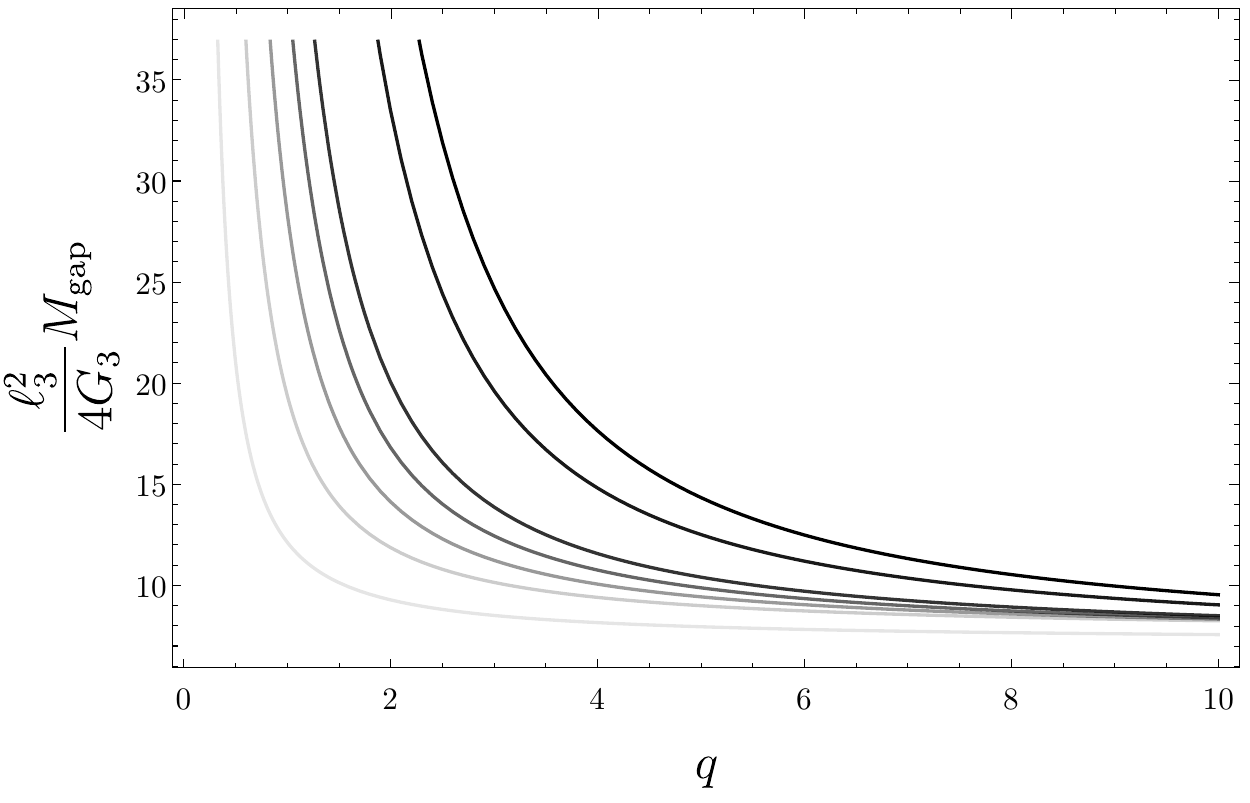}
  \end{minipage}
  \hfill
  \begin{minipage}[t]{0.45\textwidth}
    \includegraphics[width=\textwidth]{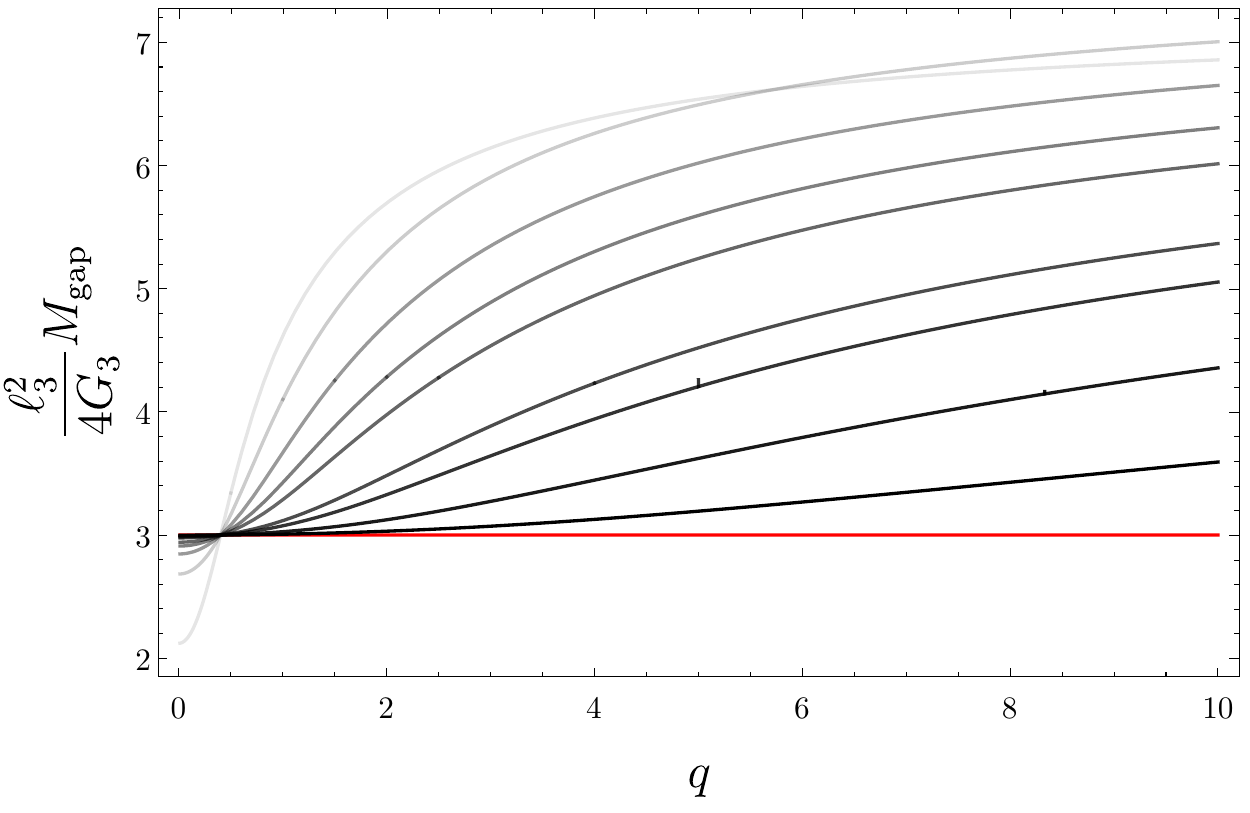}
  \end{minipage}
    \begin{minipage}[t]{0.45\textwidth}
    \includegraphics[width=\textwidth]{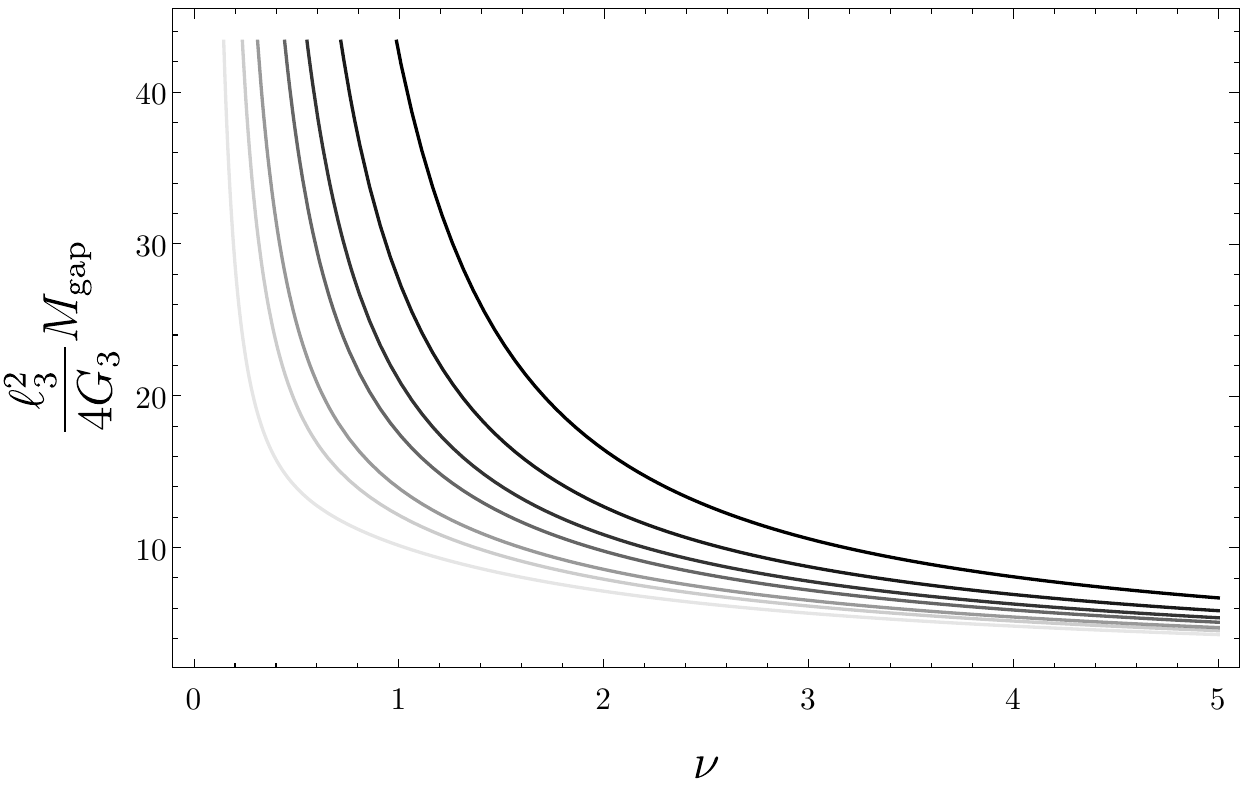}
  \end{minipage}
  \hfill
  \begin{minipage}[t]{0.45\textwidth}
    \includegraphics[width=\textwidth]{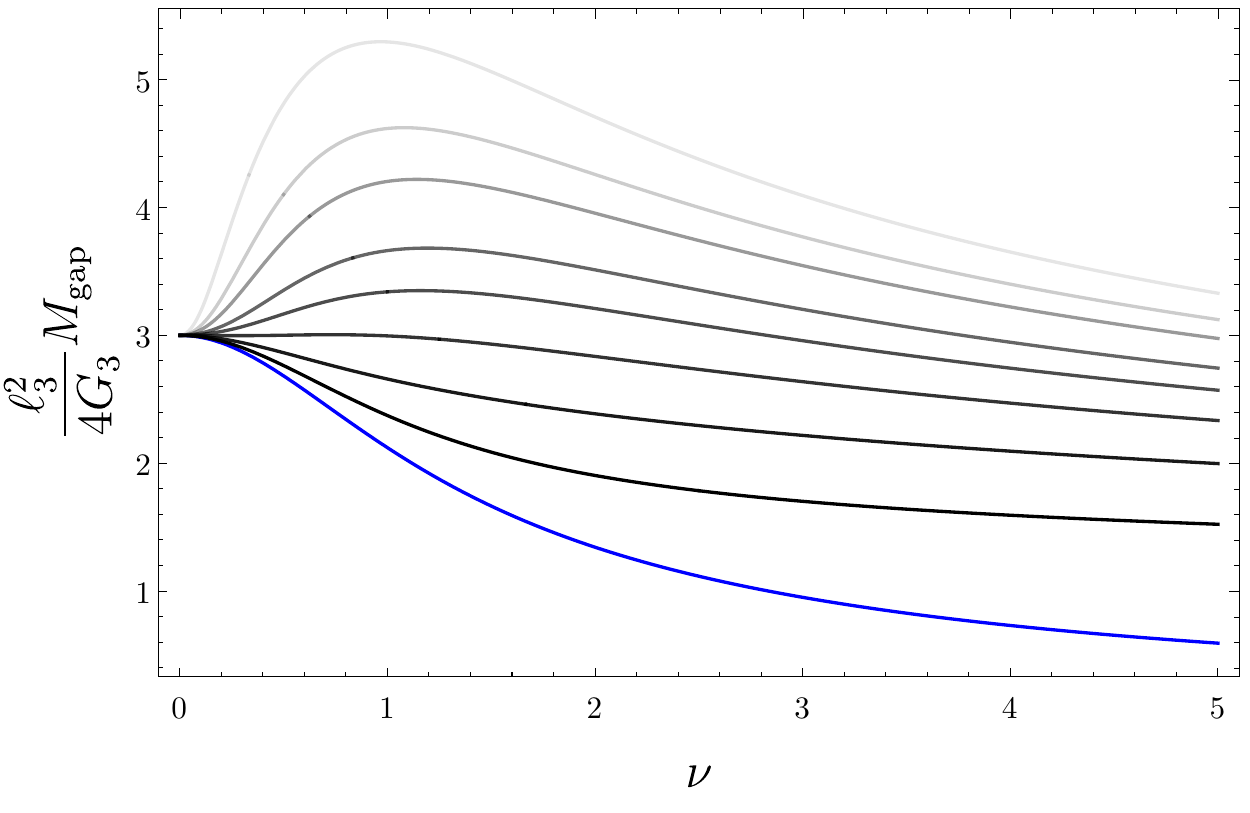}
  \end{minipage}
  \caption{\small $M_{\rm gap}$ as a function of $q$ (top) and $\nu$ (bottom) for $\kappa =+1$ (left) and $\kappa =-1$ (right). The curves with lighter colours correspond to higher values of $q$ and $\nu$. The blue and red line corresponds to the limit $q \rightarrow 0$ and $\nu \rightarrow 0$, respectively. For $\kappa=+1$, $M_{\rm gap}$ diverges in these limits, see eq.~\eqref{Mgapk1}.}
  \label{fig:Mgap}
\end{figure}

While the expression for $M_{\rm gap}$ is rather simple to state, the analysis is more subtle. This is because the parameters entering into the expression must be evaluated for the extremal solution---see appendix~\ref{App:global_aspects} for some relevant technical discussion. This information is condensed in Figure \ref{fig:Mgap}, which shows $M_{\rm gap}$ as a function of $q$ and $\nu$ for $\kappa = \pm 1$. 

First, let us discuss the case of $\kappa = +1$.  The behavior of $M_{\rm gap}$ for them is displayed in the left column of plots in Figure~\ref{fig:Mgap}. Expanding $M_{\rm gap}$ in the limit of small $\nu$ we find that 
\be \label{Mgapk1}
M_{\rm gap} = \frac{4 G_3}{\ell_3^2} \frac{1}{\nu^2 q^2}  + \dots \, ,
\ee
which indicates that quantum throat effects cannot be neglected for the black holes obtained in this limit. The reason is that the horizon of these black holes arises due to the backreaction, so when $\nu$ is small the black hole is also small and $M_{\rm gap}$ will be large.

Examining the extremal solutions we find that as the limit is approached
\be 
r_+ = \ell_3 q \nu + \mathcal{O}(\nu^3) \, , \qquad S_{\rm gen}^{\rm ext} = \frac{\pi r_+}{2 G_3} \frac{\left(1-\sqrt{1+4 q}\right)^{2}}{\left(1+ 4 q -\sqrt{1+4 q}\right) \left(1+\sqrt{1+4 q}\right)} + \mathcal{O}(\nu^3) \, .
\ee
Thus, at weak backreaction the extremal black holes are  small and their entropy approaches zero in proportion to their radius. The limiting extremal configurations we are discussing correspond to the red line in Figure~\ref{fig:backreaction}. These are the boundary defects that have $\mu = 2 q$ and form an infinitesimal extremal black hole when backreaction is turned on. In the bulk, they are similar to the extremal Reissner-Nordström black hole

The behaviour of $M_{\rm gap}$ as a function of $q$ is qualitatively similar. When $q$ is small, the expansion of $M_{\rm gap}$ has the same asymptotic dependence on $q$ as above, i.e. $M_{\rm gap} \sim q^{-2}$. These once again correspond to small extremal black holes, and quantum effects are important for this reason. On the other hand, when $q$ is large, the extremal black holes are large and quantum effects are more suppressed.

By comparing the left plot in Figure~\ref{fig:S0} with the left ones in Figure~\ref{fig:Mgap}, we see that when $\kappa=+1$ the values of $S_0$ and $M_{\rm gap}$ are inversely related to each other: both as a function of $q$ and $\nu$, when $S_0$ increases, $M_{\rm gap}$ decreases. This is generally expected: the larger the black hole is in Planck units, the smaller the scale at which quantum gravitational fluctuations are expected to appear.

Next let us consider $\kappa = -1$, described in the right column of Figure~\ref{fig:Mgap}. For small backreaction we now have
\be 
M_{\rm gap} = \frac{12 G_3}{ \ell_3^2} \left[1 + 3 q(1+q) \nu^2 + \mathcal{O}(\nu^4)\right] \, .
\ee
In the same limit, we have
\be \label{rplim}
r_+ = \frac{\ell_3}{\sqrt{3}} \left[1 + \frac{3 \nu^2 q^2}{2} + \cdots \right]
\ee
along with
\be 
S_{\rm gen}^{\rm ext} = \frac{81 \pi \ell_3 q^4 \nu^3}{8 G_3} + \mathcal{O}(\nu^5)  \, , \qquad M^{\rm ext} = \frac{19683 q^{8} \nu^{6}}{128 G_3} + \mathcal{O}(\nu^8) \, .
\ee
At first glance this behaviour may seem strange but it is actually to be expected. Recall that as $\nu \to 0$ the $\kappa = -1$ geometries limit to BTZ black holes at the boundary. These black holes do not have extremal limits with nonzero area, and therefore the limit is in this sense singular. To maintain the conditions of extremality, the geometric parameters behave as
\be 
x_1 = \frac{2 \sqrt{3}}{9 q^{2} \nu} + \mathcal{O}(\nu) \, , \qquad \Delta = \frac{81 \sqrt{3}\, q^{4} \nu^{3}}{4} + \mathcal{O}(\nu^5) \, ,
\ee
and this forces the limit to zoom in to the $M = 0$ BTZ black hole, as expected. The entropy of this black hole behaves linearly in $T$ like \eqref{nextS}, but with $S_0=0$. They have long throats but their quantum dynamics is not dominated by the Schwarzian mode. Therefore the relation between $S_0$ and $M_{\rm gap}$ for these black holes is not as straightforward as it was for the $\kappa=+1$ solutions.

On the other hand, the dependence on $q$ is now qualitatively different. At fixed $\nu$ we can expand
\be
    M_{\rm gap}  \approx \frac{4G_3}{\ell_3^2} \dfrac{3}{\sqrt{1+\nu^2}} + \mathcal{O}(q^2) \, .
\ee
This limit is also understood to be singular. At fixed $\nu$, the $q \to 0$ limit of the extremal solution would produce an extremal, negative mass hyperbolic black hole in the bulk. However, due to this metric having $\mu < 0$, the function $G(x)$ does not have any positive, real solutions in that limit, with $x_1 \to \infty$ like $x_1 \sim q^{-2}$.

\section{Final remarks}\label{sec:discuss}

Using holography we have solved several difficult problems of quantum field theory in the presence of charged sources, with or without their backreaction on the geometry. It is worth noting that even solving for a free complex scalar field in a background electric potential $A_t=a/r$ in 2+1 dimensions is not simple. This is in contrast with how a free quantum scalar in a conical geometry (a $\mathbb{Z}_n$ orbifold) or in a BTZ black hole can be readily solved through the method of images \cite{Lifschytz:1993eb,Steif:1993zv,Shiraishi:1993qnr} with results that can be usefully compared to the holographic calculation. The electric potential background cannot be solved with the same method, and thus we only have the holographic solution.

The inclusion of charge has also considerably enlarged the class of known quantum black holes in 2+1 dimensions, with the notable addition of new families of near-extremal black holes. In this article we have only obtained their main features, but a more complete analysis of the space of solutions and their physics, including quantum Schwarzian fluctuations of the throats, seems possible. Extensions to include rotation and other values of the cosmological constant on the brane can be obtained within the class of type D solutions of Einstein-Maxwell-AdS theory in \cite{Plebanski:1976gy}. This will allow to add charge to quantum black holes in $2+1$ de Sitter space of the type constructed in \cite{Emparan:2022ijy,Panella:2023lsi,future}.

\section*{Acknowledgements}

We are grateful to Antonia M.~Frassino, Sean Hartnoll, Maciej Kolanowski, Quim Llorens, Juan Pedraza, Jordi Rafecas-Ventosa, Andrew Svesko, and Manus Visser for discussions. Work supported by MICINN grants PID2019-105614GB-C22,
AGAUR grant 2017-SGR 754, PID2022-136224NB-C22 funded by MCIN/AEI/ 10.13039/501100011033/FEDER, UE, and State Research Agency of MICINN through the ‘Unit of Excellence Maria de Maeztu 2020-2023’ award to the Institute of Cosmos Sciences (CEX2019-000918-M).
The work of RAH received the support of a fellowship from ``la Caixa” Foundation (ID 100010434) and from the European Union’s Horizon 2020 research and innovation programme under the Marie Skłodowska-Curie grant agreement No 847648 under fellowship code LCF/BQ/PI21/11830027. RE is grateful to the KITP-UCSB for hospitality in the final stage of this work during the program ``What is String Theory?'', and acknowledges partial support by grant NSF PHY-2309135 and the Gordon and Betty Moore Foundation Grant No.~2919.02 to the Kavli Institute for Theoretical Physics (KITP). 

\appendix

\section{Charged Defects from Wick-Rotated Dyonic AdS Black Holes} \label{App_DW}

Let us consider the following dyonic black hole solutions in $\text{AdS}_4$ (\ref{bulk_action})
 \be \label{AdSbh}
     ds^2 = -F_k(\rho) dT^2 + \frac{d\rho^2}{F_k(\rho)} + \rho^2 d\Sigma_{k}^2
 \ee
 with 
\be \label{F_and_A}
     F_k(\rho) = \frac{\rho^2}{\ell_4^2} + k - \frac{2G_4 m}{\rho} + \frac{\tilde{Q}^2}{\rho^2} \quad \text{and} \quad A =  \frac{2}{\gl} \left[ \left(\frac{\tilde{g}}{\rho_+} - \frac{\tilde{g}}{\rho} \right) dT + \tilde{e} g_k(\theta) d\Phi \right] \,  ,
 \ee
where the $2-$dimensional metric $d\Sigma_{k}^2$ is
 \be \label{sigma_and_g}
    d\Sigma_{k}^2 =  \left\{ \begin{array}{rcl} d\theta^2 + \sin^2 \theta d\Phi^2 \quad \text{for} \ k &=& +1  \\ d\theta^2 + \sinh^2 \theta d \Phi^2 \quad \text{for} \ k &=& -1  \end{array} \right. \quad \text{and} \quad g_k(\theta) =  \left\{ \begin{array}{rcl} 1-\cos\theta \quad \text{for} \ k &=& +1 \\ s-\sinh \theta \quad \text{for} \ k &=& -1 \, . \end{array} \right.
 \ee
 In both cases, either spherical ($k=+1$) or hyperbolic ($k=-1$) black holes, $\ell_4$ is the cosmological length scale of AdS$_4$, $m$ is the ADM mass, $\tilde{Q}^2 = \tilde{g}^2 + \tilde{e}^2$ is the charge of the black hole and $\rho_+$ is the largest root of $F_k(\rho)$.
 
\subsection{AdS at the Boundary}\label{App_DW_AdS}
We can obtain a boundary geometry that is AdS$_3$ with/without conical defects. A simple way to obtain this is by starting from the hyperbolic AdS$_4$ black hole in the bulk, this is (\ref{AdSbh}), (\ref{F_and_A}) and (\ref{sigma_and_g}) with $k=-1$. 

We now carry out the double Wick rotation of this solution analogous to \cite{Hubeny:2009rc}. The coordinates and charges are transformed as
\be 
T = i \alpha \ell_4 \phi \, , \quad \Phi = i \frac{\alpha t}{\ell_3}  \, , \quad \sinh \theta = \frac{\alpha \ell_3}{R} \, , \quad \tilde{e} = - i e \, , \quad \tilde{g} = - i g \, , \quad \tilde{Q} = - i Q \, .
\ee
Here $\alpha$ is a dimensionless parameter whose value is determine by regularity of the bulk solution,
\be \label{alpha}
    \alpha = \frac{1}{\ell_4} \frac{2}{| F_{-1}'(\rho_0)|} \, .
\ee
This gives $\phi$ a period of $2\pi$. The metric and gauge field after these transformations is
\begin{align} \label{DWRN_AdS}
    ds^2 =&\, \frac{\rho^2 \alpha^2}{R^2} \left[- \left(\alpha^2 + \frac{R^2}{\ell_3^2} \right) dt^2 + \frac{dR^2}{\alpha^2 + \frac{R^2}{\ell_3^2}} + \frac{R^2 \ell_4^2}{\rho^2} F_{-1}(\rho) d \phi^2 \right] + \frac{d\rho^2}{F_{-1}(\rho)} \, ,
    \\
    A =&\, \frac{2}{\gl} \left[ - \frac{e\alpha^2}{R} dt + \frac{g\alpha \ell_4}{\rho_0} \left(1 - \frac{\rho_0}{\rho} \right) d\phi \right] \, , \label{DWRN_AdS_gauge_field} 
\end{align}
where we have set $s=0$. We can bring this metric into Fefferman-Graham form 
\be \label{FG}
    ds^2 = \frac{\ell_4^2}{z^2} \left(dz^2 + \gamma_{\mu\nu}(x,z) dx^\mu dx^\nu \right),
\ee
where the metric $\gamma_{\mu\nu}$ admits the expansion
\be 
    \gamma_{\mu\nu}(x,z) = \gamma_{\mu\nu}^{(0)}(x) + z^2 \gamma_{\mu\nu}^{(2)}(x) + z^3 \gamma_{\mu\nu}^{(3)}(x) + z^4 \gamma_{\mu\nu}^{(4)}(x) + \dots \, .
\ee

Let us perfom the following change of coordinates
\be
    \rho = \frac{\ell_4 r}{\alpha z} + \frac{\ell_4 z ( f(r) + \alpha^2)}{4 \alpha r} + \mathcal{O}(z^2) \, , \qquad R = r + \frac{f(r)}{2 r} z^2 + \mathcal{O}(z^4) \, ,
\ee
where we have introduced
\be 
    f(r) = \alpha^2 + \frac{r^2}{\ell_3^2} \, .
\ee
With this transformation, the metric $\gamma_{\mu\nu}$ takes the form:
\be
    \gamma_{\mu \nu}^{(0)} dx^\mu dx^\nu = -f(r) dt^2 + \frac{dr^2}{f(r)} + r^2 d\phi^2 \, ,
\ee
and the gauge field can be expanded as a function of $z$:
\be 
    A_t =  -\frac{2 e \alpha^2}{ \gl r} + \mathcal{O}(z^4)  \, , \qquad A_\phi = \frac{2 g \alpha \ell_4}{\gl \rho_0} \left[1 - \frac{\alpha \rho_0 z}{\ell_4 r} + \mathcal{O}(z^2) \right] \, .
\ee

\subsection{Minkowski at the Boundary}\label{App_DW_Min}
We can obtain a boundary geometry that is three-dimensional Minkowski space with or without conical defects. To do this, we once again take the hyperbolic AdS$_4$ black hole in the bulk as our starting point. This time it is convenient to use an alternate slicing for the $\mathbb{H}^2$,
\begin{align}
    ds^2 &= -F_{-1}(\rho) dT^2 + \frac{d\rho^2}{F_{-1}(\rho)} + \rho^2 \left[\frac{dR^2}{R^2} + \frac{dx^2}{R^2} \right] \, ,
    \\
    A &= \dfrac{2}{\gl} \left[ \left(\frac{\tilde{g}}{\rho_0} - \frac{\tilde{g}}{\rho} \right) dT - \frac{\tilde{e}}{R} dx \right]\, .
\end{align}
We then perform the following transformations
\be 
    T = i \alpha \ell_4 \phi \, , \qquad x = i \alpha t  \, , \qquad R \to \alpha R \, , \qquad \tilde{e} = - i e \, , \qquad \tilde{g} = - i g \, .
\ee
The resulting metric is then
\begin{align}
    ds^2 =&\, \frac{\rho^2}{R^2} \left[- dt^2 + dR^2 + \frac{\ell_4^2 F_{-1}(\rho)}{\rho^2} \alpha^2 R^2 d \phi^2 \right] + \frac{d\rho^2}{F_{-1}(\rho)} \, , \label{DWRN_Min}
    \\
    A =&\, \dfrac{2}{\ell_{\star}} \left[ - \frac{e \alpha^2}{R} dt + \alpha \ell_4 \left(\frac{g}{\rho_0} - \frac{g}{\rho} \right) d\phi \right] \, . \label{DWRN_Min_gauge_field}
\end{align}
Just as before, the parameter $\alpha$ is fixed by the regularity conditions of the geometry. In this case, it takes the form
\be 
    \alpha = \frac{2}{\ell_4} \frac{1}{|F_{-1}'(\rho_0)|} \, .
\ee
The coordinate $\phi$ in the metric is then $2 \pi$-periodic.

The metric can be brought into Fefferman-Graham form (\ref{FG}) via the transformations
\be
    \rho = \frac{\ell_4 r}{z} + \frac{\ell_4 z}{2r} + \mathcal{O}(z^2) \, , \qquad R = r + \frac{z^2}{2 r} + \mathcal{O}(z^4) \, ,
\ee
which leads to the following result for the boundary metric,
\be
    \gamma_{\mu \nu}^{(0)} dx^\mu dx^\nu = - dt^2 + dr^2 + r^2  \alpha^2 d\phi^2 \, ,
\ee
and for the gauge field 
\be
    A_t =  \dfrac{2e\alpha}{\ell_{\star}} \left[- \frac{\alpha}{r} + \frac{\alpha z^2}{2 r^3} + \mathcal{O}(z^4)  \right] \, , \qquad A_\phi = \dfrac{2g\alpha}{\ell_{\star}} \left[\frac{\ell_4}{\rho_0} - \frac{ z}{r} + \mathcal{O}(z^3)  \right] \, .
\ee

\subsection{BTZ Black Hole at the Boundary} \label{App_DW_BTZ}
We can obtain a boundary geometry that is the BTZ black hole by starting from the spherical AdS$_4$ black hole in the bulk, this is (\ref{AdSbh}), (\ref{F_and_A}) and (\ref{sigma_and_g}) with $k=+1$. 

We now carry out the double Wick rotation of this by transforming the Killing coordinates and charges as
\be 
    T = i \chi \, , \qquad \Phi = i \tau  \, , \qquad \tilde{e} = - i e \, , \qquad \tilde{g} = - i g \, .
\ee
This brings the metric and gauge field into the following form
\begin{align}
    ds^2 &= F_{+1}(\rho) d\chi^2 + \frac{d\rho^2}{F_{+1}(\rho)} + \rho^2 (d\theta^2 - \sin^2 \theta d\tau^2) \, ,
    \\
    A &= \frac{2}{\gl} \left[ \left(\frac{e}{\rho_0} - \frac{e}{\rho} \right) d\chi + g(1-\cos\theta) d\tau \right]\, .
\end{align}
We now perform one final transformation of the coordinates given by 
\be 
    \tau = \frac{r_+}{\ell_3^2} t \, , \qquad \cos \theta = \frac{r_+}{R} \, , \qquad \chi = \frac{\ell_4}{\ell_3} r_+ \phi \, ,
\ee
after which the quantities of interest take their final form
\begin{align}
    ds^2 &= \frac{\rho^2 r_+^2}{R^2 \ell_3^2} \left[- \frac{R^2-r_+^2}{\ell_3^2} dt^2 + \frac{\ell_3^2}{R^2-r_+^2} dR^2 + R^2 \frac{\ell_4^2 F_{+1}(\rho)}{\rho^2} d\phi^2 \right] + \frac{d\rho^2}{F_{+1}(\rho)} \, ,
    \\
    A &= \frac{2 e r_+}{\ell_3^2 \gl} \left(1-\frac{r_+}{R}\right) dt + \frac{2 \ell_4 r_+}{\ell_3 \gl} \left(\frac{g}{\rho_0} - \frac{g}{\rho} \right) d\phi  \, .
\end{align}
The constant $r_+$ is determined by requiring $\phi$ is $2\pi$-periodic, giving (\ref{BTZrad}). By performing a coordinate transformation, we can bring this metric into Fefferman-Graham form (\ref{FG}). We find the following change of coordinates
\be
    \rho = \frac{r \ell_3 \ell_4}{r_+  z} + \frac{\ell_4 (r^2-2 r_+^2) z}{4 r r_+ \ell_3} + \mathcal{O}(z^2) \, , \qquad R = r +  \frac{(r^2-r_+^2) z^2}{2 r \ell_3^2} +  \mathcal{O}(z^4) \, .
\ee
With this transformation, the metric $\gamma_{\mu\nu}$ admits the expansion
\be
    \gamma_{\mu\nu}^{(0)} dx^\mu dx^\nu = - f(r) dt^2 + \frac{dr^2}{f(r)} + r^2 d\phi^2 \, ,
\ee
where we have introduced
\be 
    f(r) = \frac{r^2-r_+^2}{\ell_3^2} \, ,
\ee
which is the metric function on the BTZ black hole.
Finally, we expand the components of the gauge field in the Fefferman-Graham coordinates:
\be
    A_t = \frac{2 e r_+}{\ell_3^2 \gl} \left[\left(1 - \frac{r_+}{r} \right) + \mathcal{O}(z^2) \right] \, , \qquad A_\phi = \frac{2 g r_+}{\ell_3 \gl} \left[\frac{\ell_4}{\rho_0} - \frac{r_+ z}{r \ell_3} + \mathcal{O}(z^3) \right] \, .
\ee

\section{Regularity Analysis and Parameter Ranges of Boundary Geometries}

\label{App_alpha}

In this appendix, we provide further analysis relevant to obtaining the information presented in Figure~\ref{defect_phase_space}. The main purpose is to highlight subtleties in the analysis of the different branches of solutions. For convenience, we work in units of the bulk AdS length, so $\ell_4 = 1$.

\subsection{Regular Geometries}

The boundary geometry (\ref{DWRN_AdS}) is regular when $\alpha = 1$, for which we must find a simultaneous solution to 
\be 
    F_{-1}(\rho_0) = 0  \quad \text{and} \quad |F_{-1}'(\rho_0)| = 2 \, ,
\ee
where $\rho_+$ is the largest root of $F(\rho)$. If $Q = 0$, the only consistent solution is $M = 0$, which corresponds to the original bulk geometry being AdS$_4$ written in a hyperbolic slicing. With $Q \neq 0 $ the situation is more interesting. Now, a consistent solution can be obtained corresponding to the following constraints among the parameters:
\be 
    m = 2 x^3 - x^2 - x \, , \qquad Q = x \sqrt{-3 x^2 + 2 x +1} \, ,
\ee
where $x$ is a root of $F(\rho) = 0$. While $x$ is a root of $F$, we have not called this parameter $\rho_+$ because it has not yet been determined whether $x$ is the largest root. 

\begin{figure}
\centering
\includegraphics[width=0.45\textwidth]{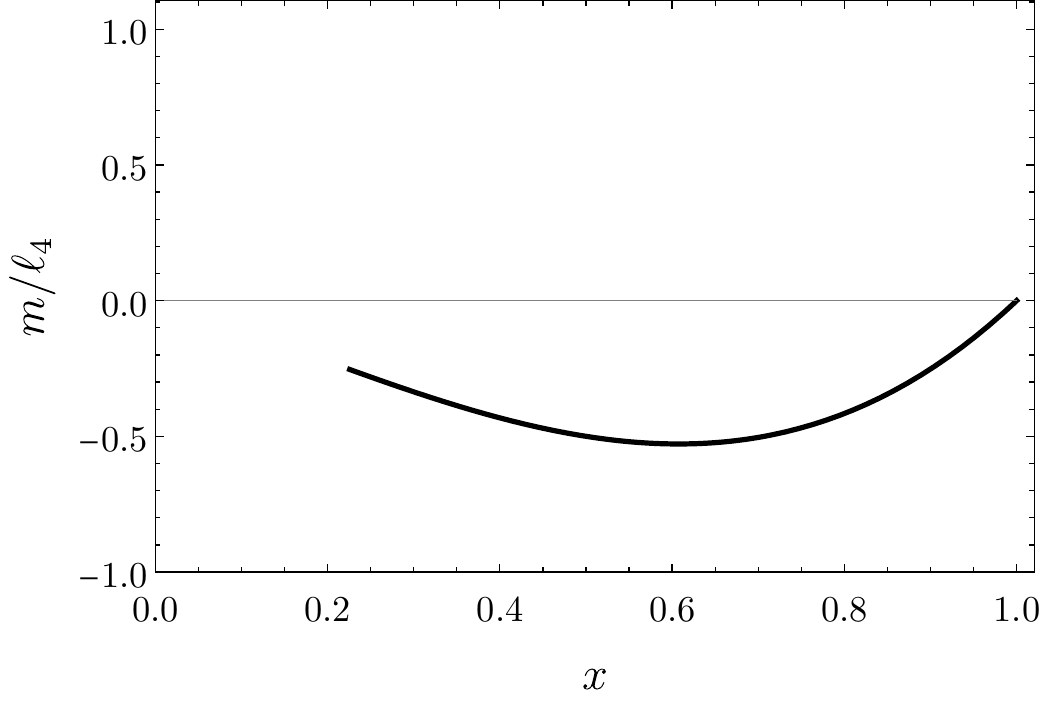}
\quad 
\includegraphics[width=0.45\textwidth]{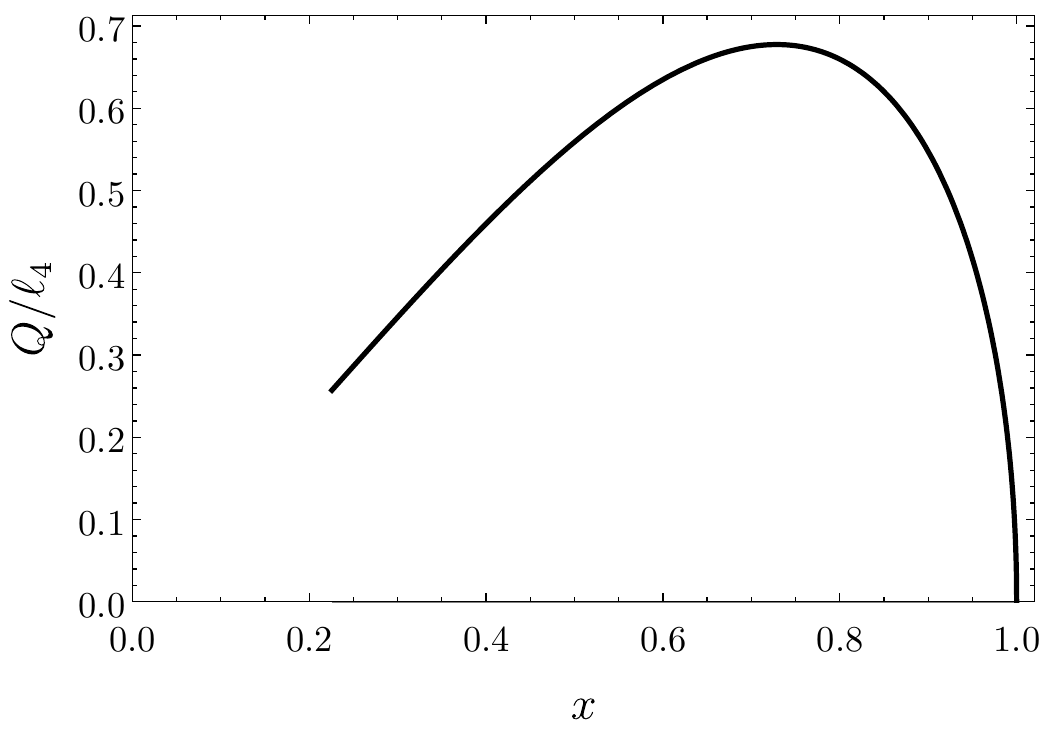}
\caption{\small The bulk mass parameter $m$ (left) and bulk charge parameter $Q$ (right) against the horizon radius. The parameters are restricted so that the boundary geometry is AdS$_3$.}
\label{regular_restrictions}
\end{figure}

\begin{figure}
\centering
\includegraphics[width=0.65\textwidth]{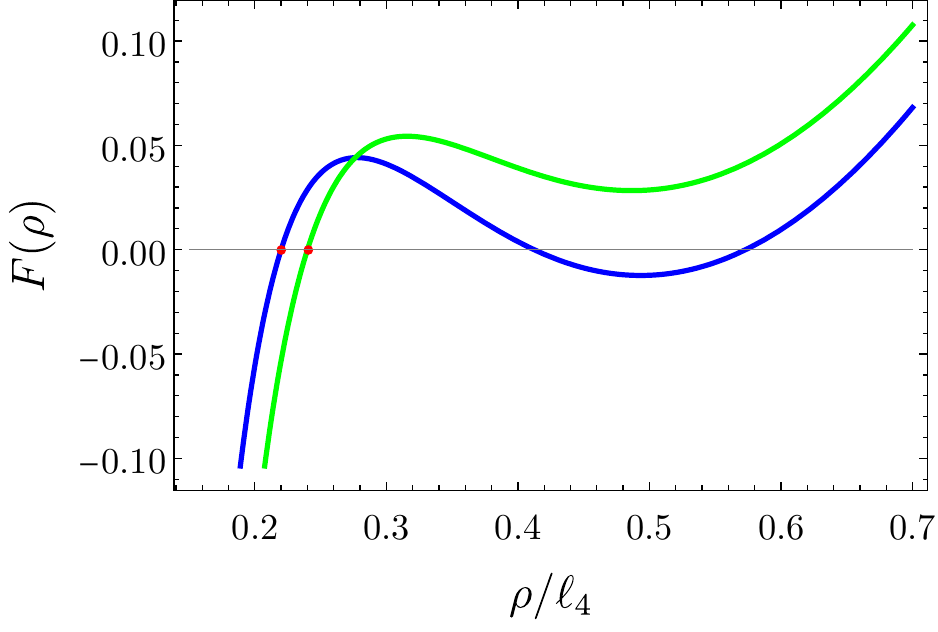}
\caption{\small The bulk metric function $F(\rho)$ for parameter $x > x_\star$ (green) $x < x_\star$ (blue).  The red dots, which are zeroes of $F$, indicate the value of $x$. As $x \to x_\star$, the curve shifts vertically downward. This shows how the value of $x$ for which the boundary solution is regular fails to be the largest root of $F$ for $x < x_\star$.}
\label{met_func_reg}
\end{figure}

We require that $Q$ is real, which constrains the allowed values of $x$ to lie in the interval $0 \le x \le 1$.\footnote{The quadratic under the square root in $Q$ has roots $-1/3$ and $1$, and is positive between these roots. Absence of naked singularities in the bulk requires positive $x$, resulting in the range given in the main text.} However, there is a stricter lower bound that arises because $\rho_+$ used in the determination of $\alpha$ must be the largest root of $F(\rho) = 0$. Requiring this further restricts the range to be
\be 
    x_\star < x \le 1 \, ,
\ee
where $x_\star \approx 0.225762$ satisfies 
\be 
    -1 + 14 x_\star^2 + 36 x_\star^3 - 33 x_\star^4 - 88 x_\star^5 + 72 x_\star^6 = 0 \, ,
\ee
which is the discriminant of $x^2F(x)$ written in terms of the quantities above.
For $x \le x_\star$ there is a root of $F(\rho) = 0$ that is larger than $x$ and regular solutions cannot be constructed. Therefore, there are bounds on the bulk parameters $(M,Q)$ over which regular boundary geometries are possible: $m_\star < m \le 0$ and $0 \le Q < Q_\star$ where
\be 
    m_\star \equiv 2 x_\star^3 - x_\star^2 - x_\star \approx  -0.253717 \, , \qquad Q_\star \equiv x_\star \sqrt{-3 x_\star^2 + 2 x_\star +1} \approx  0.257272 \, .
\ee
We illustrate this in Figure~\ref{regular_restrictions}. Further, we illustrate how exactly these bounds arise with a plot of the metric function in Figure~\ref{met_func_reg}.

\subsection{Geometries with Defects}

We now focus on the circumstance when the boundary geometry contains a charged defect in AdS$_3$ or Minkowski$_3$. 

For generic values of the bulk parameters $(m, Q)$ the boundary will contain a conical defect. First, solve the conditions 
\be 
    F_{-1}(\rho_0) = 0  \quad \text{and} \quad |F_{-1}'(\rho_0)| = \frac{2}{\alpha} \, ,
\ee
for $Q$ and $m$ as functions of $\alpha$ and $\rho_+$. Doing so gives,
\be 
    m = \frac{\rho_0(2\alpha \rho_0^2 - \rho_0 - \alpha)}{\alpha} \, , \qquad Q = \rho_0 \sqrt{\frac{- 3 \alpha \rho_0^2+ 2 \rho_0 + \alpha}{\alpha} } \, .
\ee
Now, write the metric function in terms of these reduced parameters,
\be 
    F_{-1}(\rho) = \frac{(\rho - \rho_0)\left[\alpha \rho^3 + \alpha \rho_0 \rho^2 + (1+\rho_0^2) \alpha \rho - \rho_0(3\alpha \rho_0^2 - 2 \rho_0 + \alpha) \right]}{\rho^2 \alpha} \, .
\ee
This is written in a way that makes manifest the zero of $F$ at $\rho = \rho_0$. However, the term in square brackets has a mixture of positive and negative terms so in general could have zeros. We need to examine this term to ensure it does not have a zero that is larger than $\rho_0$, which would violate the working assumption that $\rho_0$ is the largest root of $F$. 

A quick way make progress is the following. We write $\rho = \rho_+ + y$ and substitute this into the term inside the square brackets,
\be\label{threeRoots} 
    \alpha y^3 + 4 \rho_0 \alpha y^2+  (6 \rho_0^2 - 1)\alpha y + 2  \rho_0^2  \, .
\ee
If there are zeros of $F$ larger than $\rho_0$, then this polynomial will have a positive root $y$. We apply Descartes' rule of signs. All coefficients in the polynomial are positive unless $\rho_0 < 1/\sqrt{6}$, in which case the polynomial has two sign flips. Issues can arise only in this case. If $\rho_0 \le 1/\sqrt{6}$, there are either zero or two positive solutions for $y$. If it turns out there are two solutions, then there would exist a root of $F$ larger than $\rho_0$, violating the assumption that $\rho_0$ is the largest such root. This would prove that there is a smallest value of $\rho_0$, and indeed this is what we find.

To proceed further we study the discriminant of the polynomial eq.~\eqref{threeRoots}. The discriminant takes the form
\be 
    \Delta = -4 \alpha^2 \left(72 \alpha^2 \rho_0^6 - 88 \alpha \rho_0^5 - 3 (20 \alpha^2 - 9)\rho_0^4 + 36 \alpha \rho_0^3 + 14 \alpha^2 \rho_0^2 - \alpha^2 \right) \, .
\ee
We restrict our attention to the interval $0 \le \rho_+ \le 1/\sqrt{6}$. In this case, the discriminant is positive provided that 
\be\label{smallalpha} 
    \alpha < \alpha_\star \equiv \frac{2 \rho_0^3(22 \rho_0^2 - 9)  - \rho_0^2\sqrt{(3-2 \rho_0^2)^3}}{(-1+2\rho_0^2)(-1+6\rho_0^2)^2} \, , 
\ee
and otherwise negative. This gives us all the information we need. Since eq.~\eqref{threeRoots} is a cubic, it will have at most three real roots. When the discriminant is negative, there is a single real root. However, we know this must occur for negative values of $y$. This is because, as we concluded via Descartes' rule of signs, there are only ever zero or two positive roots for $y$. So, if there is a single root, it must be negative. A negative root for $y$ means that the largest root of $F$ is $\rho_0$ (recall $\rho = \rho_0 + y$) and so there is no problem. On the other hand, when $\alpha \le \alpha_\star$ the discriminant is positive. This means that all three roots of \eqref{threeRoots} are positive, and so there must be two roots of $F$ that are larger than $\rho_0$, which is a contradiction. This means there is actually a $\alpha$-dependent smallest value of $\rho_0$, and it is determined through the solution of eq.~\eqref{smallalpha}.

For a given fixed value of $\alpha$ there is also a largest value of $\rho_0$. This is determined through the requirement that $Q$ is real, which yields another (this time upper) bound on $\rho_0$.

When producing Figure~\ref{defect_phase_space}, or doing any analysis of the solutions, these bounds on the parameters must be satisfied. One cannot treat naively $\rho_0$ as a positive but otherwise free parameter.

\subsection{BTZ at the Boundary}
\label{app:BTZ_params}
\begin{figure}
    \centering
    \includegraphics[width=0.45\textwidth]{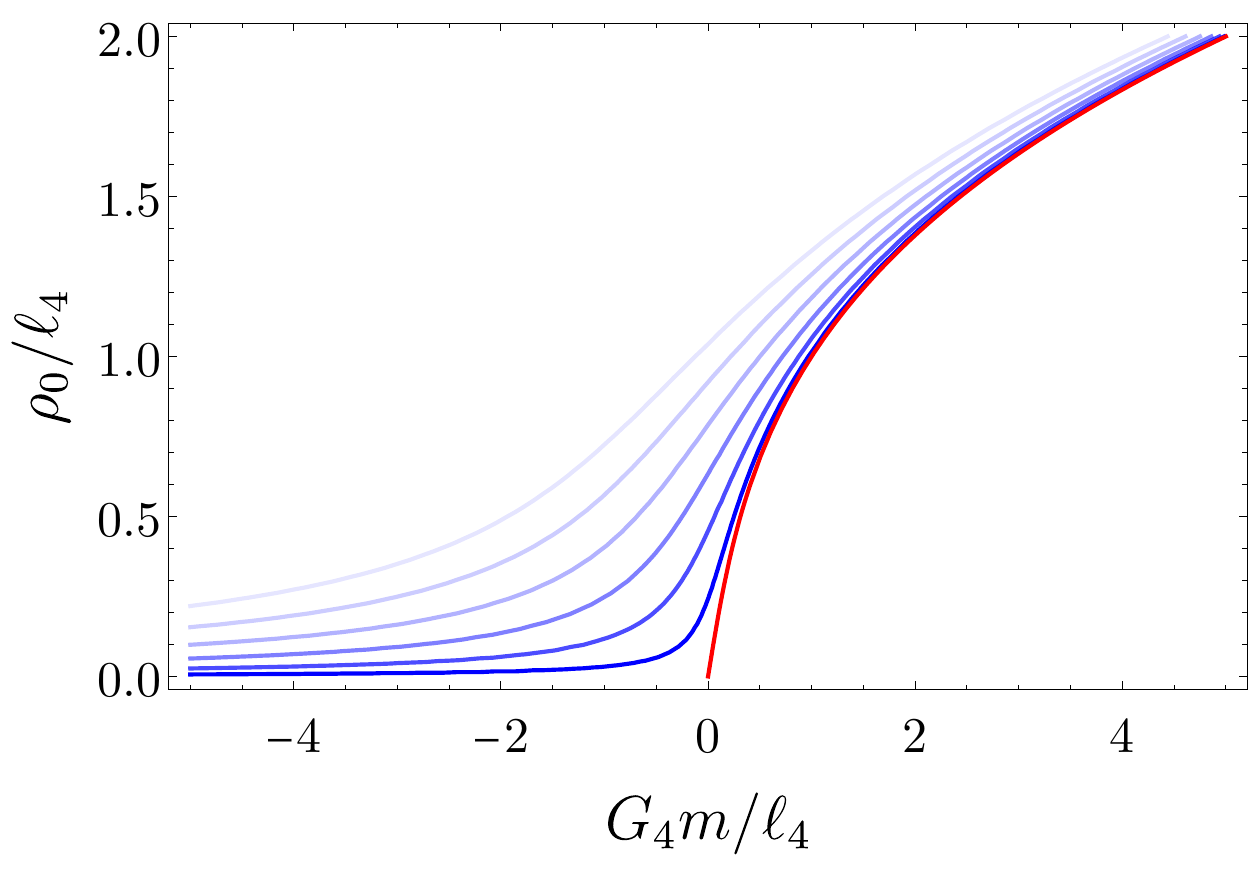}
    \quad 
    \includegraphics[width=0.45\textwidth]{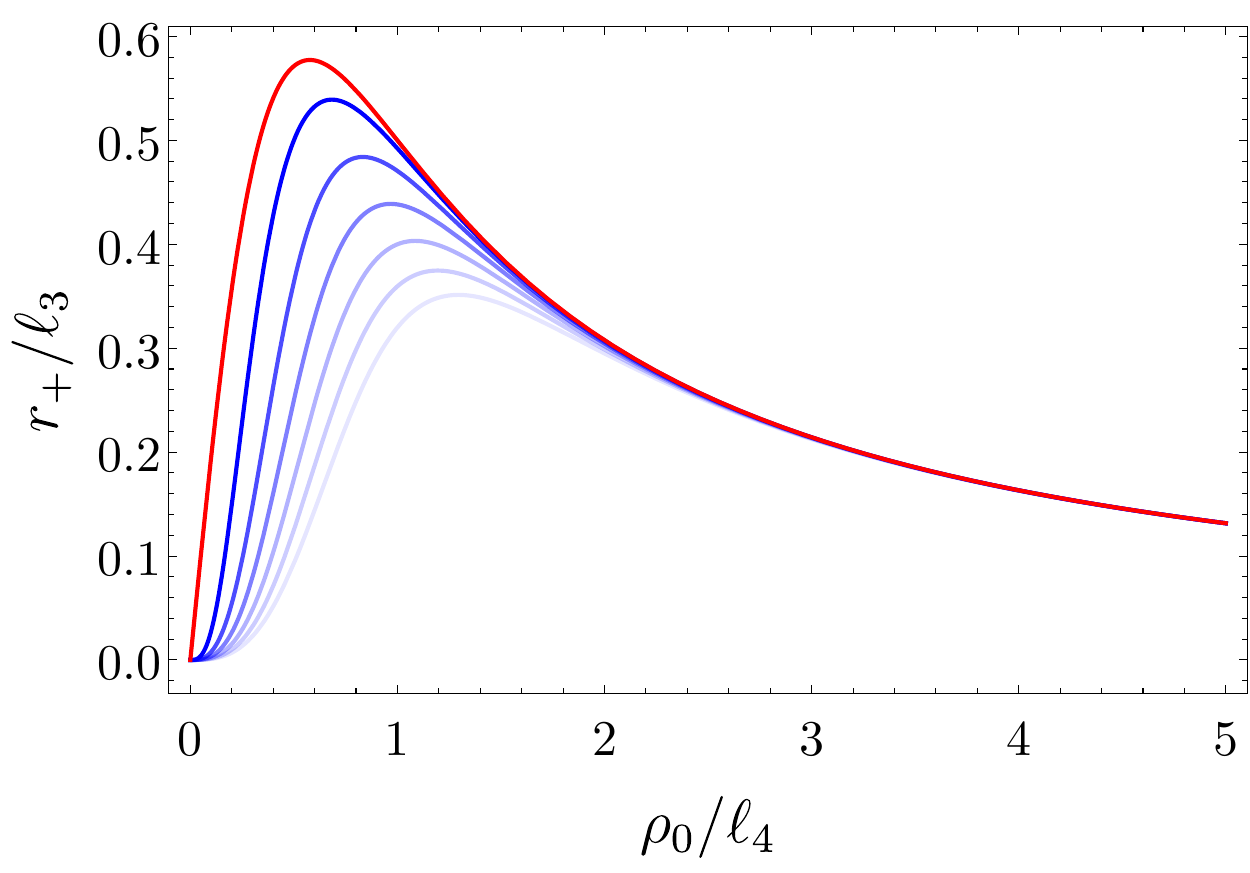}
    \caption{Left: We show the dependence of $\rho_0/\ell_4$ on the bulk parameters $(m, Q)$. Here the solid, red curve corresponds to $Q = 0$, while the blue curves correspond to different values of $Q$, with the lighter curves corresponding to larger values of $Q$. Right: We show the dependence of the BTZ horizon radius $r_+$ on the parameters $(\rho_0, Q)$. The red curve corresponds to $Q = 0$, while the blue curves have nonzero $Q$. The value of $Q$ is larger for the lighter colored curves.}
    \label{fig:btz_rad_params}
\end{figure}

Let us examine the allowed parameter ranges for setup with a BTZ black hole on the boundary. To ensure the absence of naked singularities in the bulk, we require that the metric function has a zero $F_{+1}(\rho_0) = 0$, corresponding to the constraint
\be \label{mrhoQ}
\rho_0^4 + \rho_0^2 - G_4 m \rho_0 - Q^2 = 0 \, .
\ee
Applying Descartes' rule of signs, we see that there is always a single sign flip, regardless of the sign of $m$ or $Q$. This guarantees the existence of a real, positive solution $\rho_0$ for any value of the parameter $m$, be it positive or negative. This should be constrasted with the $Q = 0$ situation, in which the existence of solutions requires $m > 0$. We show the behaviour of $\rho_0$ as a function of the solution parameters $(m, Q)$ in the left panel of Figure~\ref{fig:btz_rad_params}. Here, we see that the negative $m$ solutions correspond to smaller values of $\rho_0$, while as $m$ becomes large and positive the curves tend toward the same asymptotic behaviour as in the $Q = 0$ case. Importantly, in all cases and regardless of the value of $Q$, the parameter $\rho_0$ takes on all positive numbers, $\rho_0 \in [0, \infty)$. 

Eq.~\eqref{BTZrad} requires that the horizon radius of the BTZ black hole be given by 
\be \label{rpBTZ}
r_+ = \frac{\ell_3}{\ell_4} \frac{2}{\left| F'_{+1}(\rho_0)\right|} = \frac{2 \ell_3 \rho_0^3}{3 \rho_0^4 + \rho_0^2 + Q^2} \, .
\ee
Together with \eqref{QrelBTZ}, eqs.~\eqref{mrhoQ} and \eqref{rpBTZ} allow to translate the bulk parameters $(m,Q,\rho_0)$ into boundary data $(r_+,a,\mu_\textrm{m})$.

The analysis in this case is particularly simple. We see that $r_+ \to 0$ in the limit $\rho_0 \to 0$ and in the limit $\rho_0 \to \infty$. In the intermediate regime, the function $r_+(\rho_0)$ has a maximum at 
\be 
\rho_0 = \frac{\sqrt{6+6 \sqrt{1+36 Q^{2}}}}{6}
\ee
for which
\be 
\frac{r_+^{\rm max}}{\ell_3} = \frac{\left(1+\sqrt{36 Q^{2}+1}\right) \sqrt{6+6 \sqrt{36 Q^{2}+1}}}{72 Q^{2}+6 \sqrt{36 Q^{2}+1}+6} \, .
\ee
The essential detail is that the maximum value of $r_+$ is never larger than the maximum that occurs for $Q = 0$. Since the temperature of the black hole has a simple relationship with the horizon radius, eq.~\eqref{TBTZ},
these observations directly translate into a statement about the temperature. In particular, we can conclude that, for any fixed $Q$, there is a maximum temperature and this maximum temperature decreases as the charge of the defect increases. We show  $r_+$ as a function of bulk parameters in the right panel of Figure~\ref{fig:btz_rad_params}, illustrating the behaviour we have just described.

\section{Global Aspects of the Bulk and the Brane} 
\label{App:global_aspects}

\subsection{Range of Parameters in the Bulk and Extremality}\label{app:extremal}

Let us now make some general remarks about the global structure. The existence of horizons is governed by the metric function $H(r)$, while $G(x)$ determines the geometry of the transverse sections. Therefore, the root structure of $G(x)$, and in particular the allowed values of the smallest root $x_1$ as a function of the solution parameters, determines the possible solutions that can be described on the brane. For ease of comparison with the quBTZ case, we will use $x_1$ and $q$ as primary parameters with 
\be \label{app_mu_x1}
    G(x_1) = 1-\kappa x_1^2 - \mu x_1^3 - q^2x_1^4 = 0 \quad \Rightarrow \quad \mu = \dfrac{1-\kappa x_1^2- q^2x_1^4}{x_1^3} \, .
\ee
We will need to determine the allowed range of $x_1$ as a function of $q$. In order to do it we implement only two requirements: (a) the bulk geometry should have a horizon $r_0 > 0$, (b) the surfaces of constant bulk $(t,r)$ should be compact, i.e.~$x_1$ should be generally a finite number. In the case of $\kappa = 0, +1$ the requirement of horizons necessitates $\mu > 0$. Meanwhile, for $\kappa = -1$ it is possible for horizons to exist for both positive and some range of negative $\mu$. We find that the key difference here is that extremal black holes are possible in the bulk for all values of $\kappa$ which result in a bound on the allowed values of $\mu$ in all cases. In eq.~\eqref{muqr0} we gave the extremal values of $\mu(q)$ in parametric form $\left(\mu(r_0),q(r_0)\right)$, with $r_0$ the horizon radius. This is conveniently simple but it does not fully discriminate among branches corresponding to square root sign choices. Therefore we give here the explicit forms
\be 
    \mu_{\rm ext}^{(\kappa)} = \sqrt{\frac{2}{3}}  \frac{\sqrt{\sqrt{\kappa ^2+12 \nu ^2 q^2}-\kappa } \left(2 \kappa +\sqrt{\kappa ^2+12 \nu ^2 q^2}\right)}{3 \nu } \, , \qquad r_0^{(\kappa)} = \frac{\ell_3 \sqrt{\sqrt{\kappa ^2+12 \nu ^2 q^2}-\kappa }}{\sqrt{6}} \, .
\ee
When $\mu = \mu_{\rm ext}^{(\kappa)}$ the maximum value of $x_1$---$x_1^{\rm max}$---is obtained. Recall that in the cases where the braneworld holography can be trusted $\nu$ should be a small parameter. We can expand $\mu_{\rm ext}^{(\kappa)}$ for small $\nu$ and see the following behaviours:
\begin{align}
    \mu_{\rm ext}^{(+1)} &= 2q + \nu^2 q^3 + \mathcal{O}(\nu^4) \, , 
    \\
    \mu_{\rm ext}^{(0)} &= \frac{4 \sqrt{\nu } q^{3/2}}{3^{3/4}} + \mathcal{O}(\nu^{7/2}) \, ,
    \\
    \mu_{\rm ext}^{(-1)} &= -\frac{2}{3 \sqrt{3} \nu } +  \sqrt{3} \nu  q^2 + \mathcal{O}(\nu^{3})  \, .
\end{align}
What is notable here is that for $\kappa = -1$ the extremal value of $\mu$ depends on an inverse power of $\nu$, and hence should generally be very large. It is also negative in this case.

To determine the value of $x_1$ in the extremal case we must solve $G(x_1) = 0$ with $\mu = \mu_{\rm ext}^{(\kappa)}$. This can be done analytically, but to gain some insight it is useful to look at perturbative expansions at small and large values of $q$ and also considering small values of $\nu$. In the case of large $q$ we get
\be
    x_1^{\rm max} = \frac{1}{\sqrt{q}} \left[ h(\nu) - \frac{\kappa}{4 \times 3^{1/4}} \frac{1}{\sqrt{\nu} q} + \frac{\kappa^2 \left(3^{1/4} h(\nu) + 2 \sqrt{3 \nu} \right)}{96 h(\nu) \nu^{3/2}} \frac{1}{q^2} + \cdots \right] \, ,
\ee
with 
\be \label{hnu}
    h(\nu) = 1 - \frac{\sqrt{\nu}}{3^{3/4}} + \frac{\nu}{2 \sqrt{3}} + \dots \, .
\ee

Let us now consider the opposite limit when $q$ is small, we find 
\begin{align}
    x_1^{\rm max} &= \frac{h(\nu)}{\sqrt{q}}  \quad \text{for } \kappa = 0 \, ,
    \\
    x_1^{\rm max} &= 1 - q + 2 q^2 + \dots  \quad \text{for } \kappa = +1 \, ,
    \\
    x_1^{\rm max} &= \frac{2}{3 \sqrt{3} \nu q^2}  + \frac{\sqrt{3} \nu}{2} - \frac{21 \sqrt{3} \nu^3 q^2}{8} + \dots  \quad \text{for } \kappa = -1 \, .
\end{align}
We see that for $\kappa=0$ the full and exact solution is $x_1^{\rm max} = \frac{h(\nu)}{\sqrt{q}}$. However, for $\kappa=\pm 1$ we conclude that in all cases we recover the appropriate quBTZ result for $x_1^{\rm max}$ as $q \to 0$. Again here we note that for $\kappa = -1$ the result depends on $1/\nu$, which should be a large number in the domain of validity of braneworld holography.\\ 

\subsection{Bounded Mass of the Charged quBTZ Black Hole}
\begin{figure}[t]
  \centering
  \begin{minipage}[t]{0.45\textwidth}
    \includegraphics[width=\textwidth]{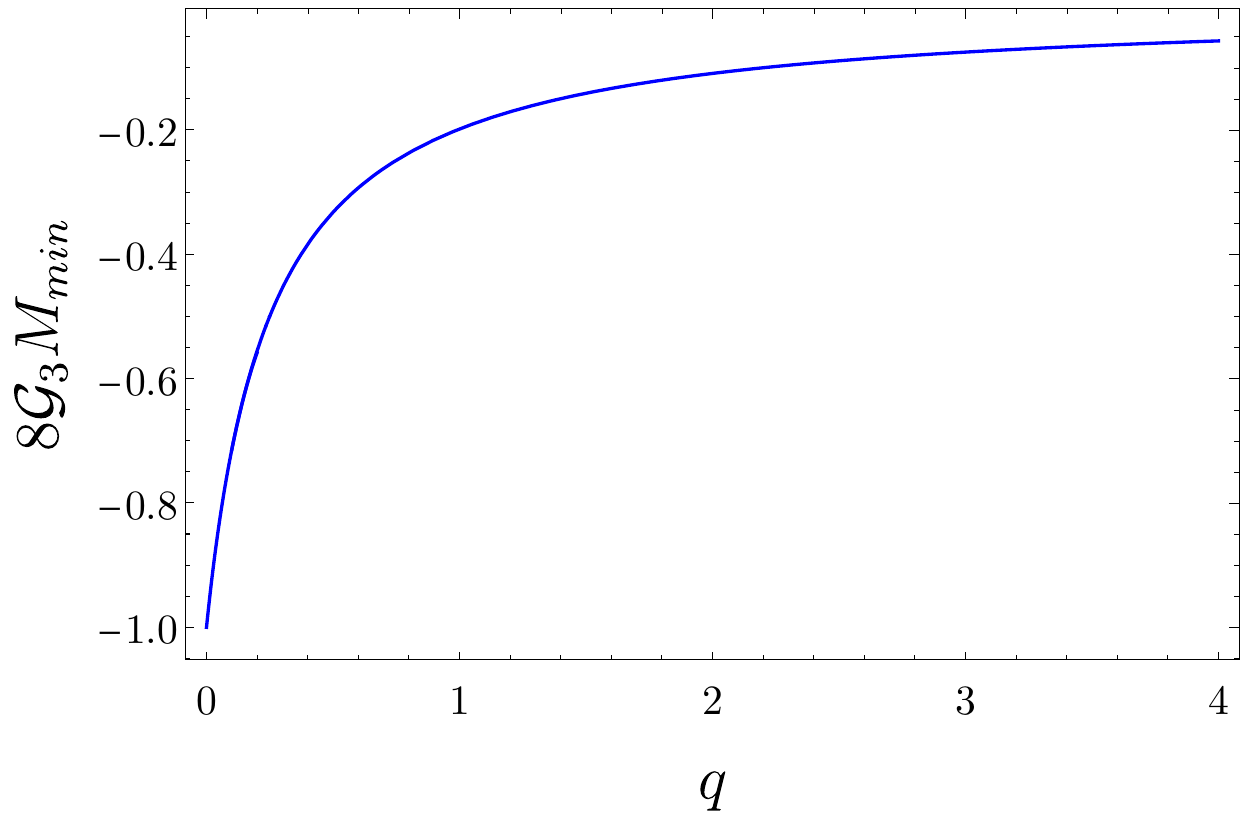}
  \end{minipage}
  \hfill
  \begin{minipage}[t]{0.45\textwidth}
    \includegraphics[width=\textwidth]{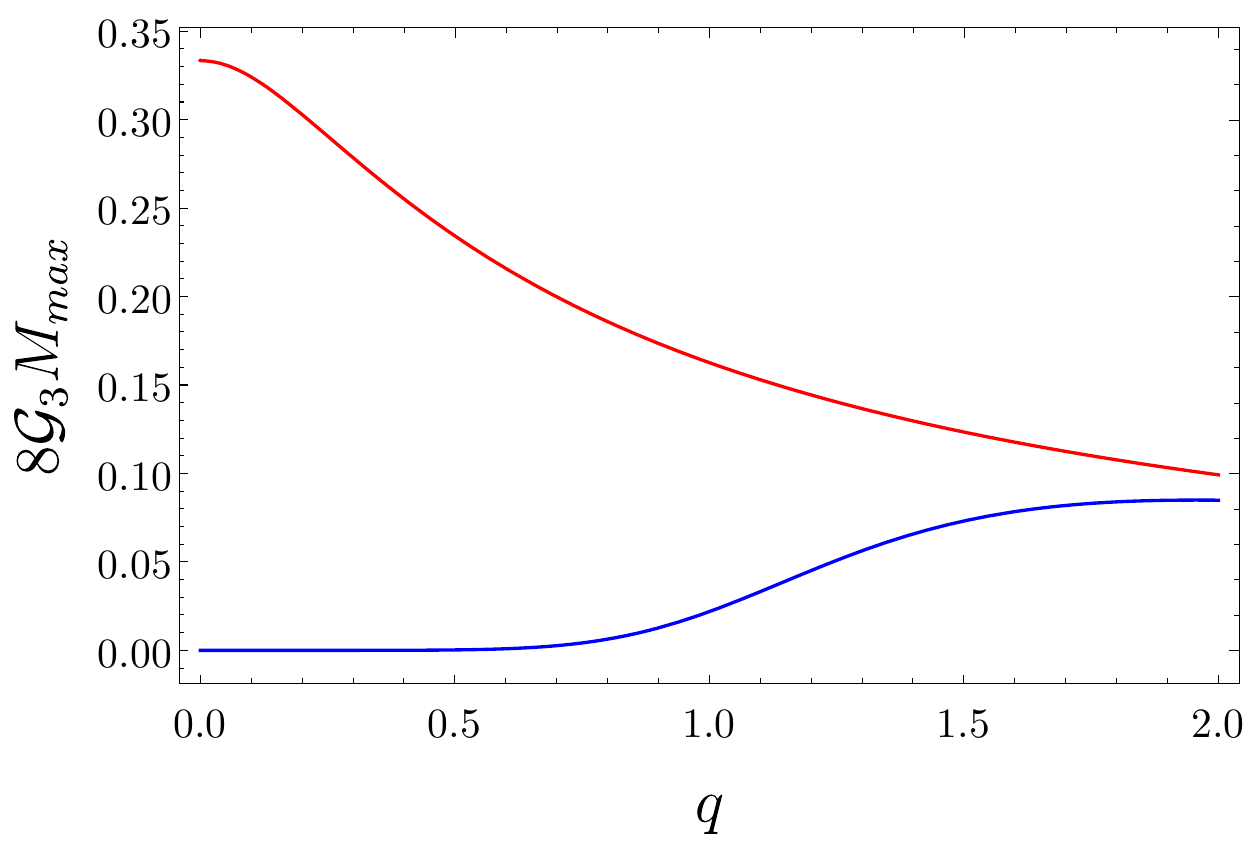}
  \end{minipage}
  \caption{\small Left: Minimum mass of the charged qBTZ as a function of $q$ for $\nu=1/5$. It corresponds to an extremal black hole. Right: Maximum mass of the charged qBTZ as a function of $q$ for $\nu=1/5$. The red line is the maximum mass of the black hole $M_{\text{max}}$ allowed by the bulk. The blue line is the extremal limit.}
  \label{Fig:mass_min_max}
\end{figure}

The range of masses (\ref{M_F}) is always bounded, and in fact a proper subset of the allowed masses for the quBTZ black hole. In this case, however, the minimum and maximum allowed masses is a function of both $q$ and $\nu$. The minimal mass is achieved for the $\kappa = +1$ black holes, and coincides with the extremal limit in all cases (except for $q = 0$, in which case the minimum mass corresponds to global AdS$_3$). Let us explore the range of masses (\ref{M_F}) of the charged quantum black holes of maximum mass. As a function of $x_1$, $M$ will always have a maximum. This occurs when
\be 
    \frac{dM}{dx_1} = 0 \Rightarrow x_1^{\rm turn} = \frac{\sqrt{-1 + \sqrt{1+36 q^2}}}{\sqrt{6} q} \, .
\ee
If this value of $x_1^{\rm turn}$ is less than $x_1^{\rm max}$ for the particular values of $q$ and $\nu$, then the maximum of the mass will occur at this point, and will be given by
\be 
    M_{\rm max} = \frac{27 q^2 \left(\sqrt{36 q^2+1}-1\right)}{4 \left(36 q^2+\sqrt{36 q^2+1}-1\right)^2} \, .
\ee
However, it is also possible that $x_1^{\rm max} < x_1^{\rm turn}$. If this is the case, then it is the extremal black hole that will have maximum mass---see Figure \ref{Fig:mass_min_max}.\\

\bibliography{bib}

\end{document}